# Structure, spectroscopy and cold collisions of the $(SrNa)^+$ ionic system

**Sana Bellaouini[1], Arpita Pal[2], Arpita Rakshit[3], Mohamed Farjallah[1], Bimalendu Deb[2,4], Hamid Berriche[1,5,*]**

[1]*Laboratory of Interfaces and Advanced Materials, Physics department, Faculty of Science, University of Monastir, 5019 Monastir, Tunisia.*

[2]*Department of Materials Science, Indian Association for the Cultivation of Science, Jadavpur, Kolkata 700032, India*

[3]*Sidhu Kanhu Birsa Polytechnic, Keshiary, Paschim Medinipur 721133, India.*

[4]*Raman Center for Atomic, Molecular and Optical Sciences, Indian Association for the Cultivation of Science, Jadavpur, Kolkata 700032, India.*

[5]*Mathematics and Natural Sciences Department, School of Arts and Sciences, American University of Ras Al Khaimah, Ras Al Khaimah, PO Box 10021, RAK, UAE.*

**Corresponding author: hamid.berriche@aurak.ac.ae, hamid.berriche@fsm.rnu.tn*

# Abstract

We perform a study on extended adiabatic potential energy curves of nearly 38 states of $^{1,3}\Sigma^{+}$, $^{1,3}\Pi$ and $^{1,3}\Delta$ symmetries for the $(SrNa)^{+}$ ion, though only the ground and first two excited states are used for the study of scattering processes. Full Interaction Configuration (CI) calculations are carried out for this molecule using the pseudopotential approach. In this context, it is considered that two active electrons interact with the ionic cores and all single and double excitations were included in the CI calculations. A correction including the core-core electron interactions is also considered. Using the accurate potential energy data, the ground state scattering wave functions and cross sections are obtained for a wide range of energies. We find that, in order to get convergent results for the total scattering cross sections for energies of the order 1 K, one need to take into account at least 87 partial waves. In the low energy limit ( < 1 mK), elastic scattering cross sections exhibit Wigner law threshold law behavior while in the high energy limit the cross sections go as $E^{-1/3}$. A qualitative discussion about the possibility of forming the cold molecular ion by photoassociative spectroscopy is presented.

**Keywords:** Ab initio, Adiabatic potential energy, vibrational levels, spectroscopic properties, dipole moment, cold collision, and photoassociation.

## I. Introduction

Over the years, a great number of scientific explorations have taken place on the basis of particle-particle scattering establishing the fundamental nature of particles. With the development of laser cooling and trapping of neutral and ionic atoms, the translational motion of particles can now be slowed to the point where the effects of quantum statistics begin to affect the overall behavior of the gas, setting a new era in physics of cold particles has started. It is now possible for atoms to interact with one another for a long time during a collision event at an extremely low temperature. This allows one to investigate and understand the weak long-range forces that atoms exert on one another. Ion-atom scattering at low energy is important for a number of physical systems, such as cold plasmas, planetary atmospheres, interstellar clouds, etc. At ultralow temperatures, atom-atom and atom-ion scattering processes under central potentials converge to the limit of the lowest quantum mechanical partial wave, in which the dynamics of the system can usually be described in terms of the *s*- wave scattering length. Low temperature condition greatly simplifies the problem since it freezes both rotational and vibrational dynamics and reduces the number of contributing partial waves to a low-enough value that one can perform true quantum scattering calculations. Relying on these properties, one can investigate the possibilities of forming cold molecular ion by photoassociative spectroscopy and understand the laser-induced cold chemical reactions between atoms and ions. Gaining insight into the ion-atom interactions and scattering at ultra-low energy down to Wigner threshold law regime is important to understand charge transport [1] at low temperature, quantum reaction dynamics, polar ion physics [2], ion-atom bound states [3], ion-atom photoassociation [4], etc. Several theoretical as well as experimental studies [5] of atom-ion cold collisions have been carried out in recent times. It is proposed that controlled ion-atom cold collision can be utilized for quantum information processes [6].

Here we study cold collision of ion-atom systems. Our case is for alkali ion ($A^+$) and alkaline earth metal atom (B). Atomic mass of B is greater than that of A. The ground state $X^1\Sigma^+$ corresponds to $A^{(+)}B(ns)$. The colliding atom-ion pair we consider is $(SrNa)^+$. To calculate scattering wave functions, we need the data for Born-Oppenheimer adiabatic potentials of the systems. Therefore, accurate potentials for the ground and excited state are needed. Many excited states are extensively studied to determine their spectroscopic constants, permanent and transition dipole moments. Large basis sets are optimized for a better representation of the atomic energy levels of $Sr^+$, Sr and Na.

In this paper, we present detailed results of potential energies, spectroscopic properties, dipole moments and scattering cross sections over a wide range of energies. The paper is organized in the following way.

## II. Structure and spectroscopy of $Sr^+Na$

### II. 1. Method of calculation

The use of the pseudopotential method, for $Sr^{2+}$ and $Na^+$ cores in the $SrNa^+$ ionic molecule, reduce the number of active electrons to only one electron. We have used a core polarization potential $V_{CPP}$ for the simulation of the interaction between the polarizable $Sr^{++}$ and $Na^+$ cores with the valence electron. This core polarization potential is used according to the formulation of Muller et al [7]:

$$V_{CPP} = -\frac{1}{2}\sum_{\lambda} \alpha_{\lambda} \vec{f}_{\lambda} \cdot \vec{f}_{\lambda}$$

where $\alpha_{\lambda}$ represents the dipole polarizability of the core $\lambda$ and $\vec{f}_{\lambda}$ represents the electric field created by valence electrons and all other cores on the core $\lambda$.

$$\vec{f}_{\lambda} = \sum_{i} \frac{\vec{r}_{i\lambda}}{r_{i\lambda}^{3}} F(\vec{r}_{i\lambda}, \rho_{\lambda}) - \sum_{\lambda' \neq \lambda} \frac{\vec{R}_{\lambda'\lambda}}{R_{\lambda'\lambda}^{3}} Z_{\lambda}$$

$F_l$ represents the dependent cutoff function, following the Foucroult et al [8] formalism, where $\vec{r}_{i\lambda}$ is a core-electron vector and $\vec{R}_{\lambda'\lambda}$ is a core-core vector.

Furthermore, an l-adjustable cut off radius has been optimized to reproduce the atomic ionization and the lowest energy levels: s, p, d and f. They are presented in tables 1 and 2. For the Sr atom, we used a relatively large uncontracted Gaussian basis set 5s5p and 5s4d, which is sufficient to reproduce the involved neutral and ionic atomic levels. While for the Na atom, we used a (7s5p7d2f/6s5p5d 2f) GTOs basis set. We have considerably extended the used basis sets to permit a much wider exploration, particularly for the higher excited states. The used core polarizabilities for the $Na^+$ and $Sr^{2+}$ cores are, respectively, 0.993 and 5.51 $a^3_0$. Using the pseudopotential technique the ionic molecule $(SrNa)^+$ is reduced to only two valence electrons interacting with two cores, $Sr^{2+}$ and $Na^+$. Within the Born-Oppenheimer approximation an SCF and a full CI calculations, provide us with very accurate potential energy curves and dipole functions. The calculation have been performed using the CIPSI package (configuration interaction by perturbation of multiconfiguration wavefunction selected iteratively) developed at the Laboratoire de Physique et Chimie Quantique of Toulouse in France.

**II. 2. Potential energy curve and spectroscopic constants**

We display in figures 1-6 the potential energy curves (PECs) for $^{1,3}\Sigma^+$, $^{1,3}\Pi$ and $^{1,3}\Delta$ symmetries below $Sr^{2+}Na^-$ dissociating limit of the $SrNa^+$ ionic molecular system. The calculation was carried out for a wide range of internuclear distances varying from 3.5 to 150 a.u with a step of 0.05 a.u around the avoided crossings, in order to cover all the ionic-neutral crossings in the different symmetries. This is particularly important for the adiabatic representation, around weakly avoided crossings. The description and examination becomes more complicated for the higher states, which are closed to each other and interact much more for various bond distances. Noteworthy, an interesting behavior can be observed for the excited states of $^{1,3}\Sigma^+$ symmetries. Series of undulations can be seen in Fig 1–2, which lead to potentials with double and sometimes triple wells. Some states, especially the lowest ones, are smooth with a well-defined unique minimum. Series of avoided crossings related to charge transfer can be clearly seen. For example the $3^1\Sigma^+$ and $4^1\Sigma^+$ states interact on two occasions 13.68 a.u and 18.18 a.u. As it can be seen from the figures most states are bound and present potential depths varying from hundreds to thousands $cm^{-1}$. Aymar et al. [9-10] reported the first calculation for the (X, A, B and C) $^{1,3}\Sigma$, (1 and 2)$^1\Pi$, (1, 2 and 3)$^3\Pi$ and (1and 2)$^3\Delta$ and $1^1\Delta$ electronic states. Recently Ben Hadj Ayed et al [11] reported extended calculations using the same calculation techniques and covering more excited states than Aymar et al.

Generally, the knowledge of the spectroscopic constants guide experimentalists especially in the identification of new molecular species. In order to check the precision of our potential energy curves for the states previously studied, we have extracted the spectroscopic constants of the ground and low-lying states, which are reported in Table 3 and 4. They are compared with the available theoretical works of Aymar et al [9-10] and Ben Hadj Ayed et al [11]. The reported spectroscopic constants are: the equilibrium distance $R_e$, the well depth $D_e$, the electronic excitation energy $T_e$, harmonic frequency $\omega_e$, the anharmonicity constant $\omega_e x_e$ and the rotational constant $B_e$. Our results are in general agreement with those of Aymar et al and Ben Hadj Ayed et al. For the ground state, we obtained values in excellent agreement with their spectroscopic constants; especially for the equilibrium distance $R_e$= 6.92 a.u, and for the potential well depth $D_e$ = 8408 $cm^{-1}$. These values are compared with those of Aymar et al ($R_e$=6.9 a.u and $D_e$= 8499 $cm^{-1}$) and those of Ben Hadj Ayed et al ($R_e$= 6.85 a.u, $D_e$= 8361.57 $cm^{-1}$). That is also the case for A, B and C $^1\Sigma^+$ excited states. This is not surprising since we use a similar technique. The comparison of our equilibrium distance and the dissociation energy with the molecular constants given by Aymar et al for the $1^3\Sigma^+$ excited state shows a good accord with relative differences of $\Delta(R_e)$~0.00%, $\Delta(\omega_e)$~0.02%, and $\Delta(D_e)$ ~ 0.42%. However, Ben Hadj Ayed et al reported for this state two minimums close to each other. Their second well can be identified to the only one we and Aymar et al found. Our equilibrium distance of 8.03 a.u is the same that Aymar et al found. This same excellent agreement is observed for $D_e$ and $w_e$. The first and absolute minimum of 7.871 a.u identified by Ben Hadj Ayed et al for this state is not found neither in our calculation nor in that of Aymar et al. In addition, Ben Hadj Ayed et al reported a very close harmonic frequency ($\omega_e$=120.82 $cm^{-1}$) compared to our ($\omega_e$=118.02 $cm^{-1}$) and that of Aymar et al ($\omega_e$ = 121 $cm^{-1}$).

The 2 $^3\Delta$ excited state is found by Aymar et al. to be attractive with a potential well of 134 $cm^{-1}$ located at an equilibrium distance of 17.8 a. u and a vibrational frequency of 35 $cm^{-1}$. Although we found a close well depth, $D_e$=152 $cm^{-1}$, the equilibrium distance and vibrational frequency are underestimated in our work, $R_e$=13.09 a. u and *we*=13.09 $cm^{-1}$. Regarding the $^{1,3}\Pi$ symmetry, the potential energy curves of the (1-6) $^{1,3}\Pi$ states are drawn in figures 3 and 4 as function of separations R. We note that these PECs present usual and regular shapes. The electronic states of $^{1,3}\Pi$ symmetries are characterized by single shallow wells except the 3, 6 $^1\Pi$ states, which have double wells at 9.44, 14.71, and 9.70, 19.43 a.u., respectively. While, the 4 $^1\Pi$ electronic state is repulsive. Series of undulations can be visibly observed at various separations R leading to potentials with doublet wells. Therefore, we see the presence of series of avoided crossings related to charge transfer between the electronic states. To our knowledge, there are not experimental results available for these main spectroscopic constants (equilibrium distance $R_e$ and potential well depth $D_e$) and definitive conclusions cannot be drawn.

To conclude, a general agreement is observed between the three calculations, which is not surprising as same techniques were used based on pseudopotentials and Interaction of Configurations. However, a better agreement is seen between our work and that of Aymar et al [9-10] for most of the electronic states. Detailed comparisons are presented in tables 3 and 4.

## II. 3. Permanent and transition dipole moments

Comprehension of the dynamics of diatomic molecules at ultralow temperatures can be manipulated and controlled with the external static electric or non-resonant laser fields which couple with the permanent electric dipole moment and electric dipole polarizability, respectively [12,13] and in order to realize absorption measurements of the rotational transitions the permanent dipole moment must exist. We note that the dipole moment is one of the most important parameters that determines the electric and optical properties of molecules. For these simple reasons, we calculated the adiabatic permanent dipole moments for a large and dense grid of internuclear distances. To understand the ionic behavior of the excited electronic states, we have presented in Figures 7-11 the permanent dipole moment of the $^1\Sigma^+$, $^3\Sigma^+$, $^1\Pi$, $^3\Pi$, $^1\Delta$ and $^3\Delta$ symmetries states function of the internuclear distance. Before discussing in detail our results on permanent and transition dipole moments for the $SrNa^+$ ionic molecular system, we mention that most of our results were not available elsewhere; the only available study on dipole moment for the $SrNa^+$ ionic molecule, is published by Aymar et al. [9-10] for the ground state $X^1\Sigma^+$. This lack of results for $SrNa^+$ and for the similar systems which have alike electronic properties encouraged us to make more efforts to determine new information especially for the excited states for $^1\Sigma^+$ symmetry and for other symmetry. This will guide experimentalists and theorists who have similar interests about formation prediction of ionic alkali earth-alkali diatomic molecules. We start our discussion by the PDM (Permanent Dipole Moment). It is clear in figures 7-11 that the significant changes of the sign of the permanent dipole moment at small internuclear distances are due to change of the polarity in the molecule, going from the $SrNa^+$ structure for the positive sign to the $Sr^+Na$ structure for the negative sign. In addition, we remark the presence of many abrupt changes in the permanent dipole moments. This is very clear in Figure 7, which shows the permanent dipole moment of the first ten states of $^1\Sigma^+$ symmetry. We remark that, for the short internuclear distances, the adiabatic permanent dipole moments vary smoothly and sometimes they exhibit some abrupt variations such as that between $9^1\Sigma^+$/ $10^1\Sigma^+$ and between $3^1\Sigma^+$/$4^1\Sigma^+$ states which occurs respectively at 26.32 a. u. at 16.47 a. u This appears clearly in the zoom for the short distances. We remind that these particular distances represent position of the avoided crossing between the two adiabatic potential energy curves ($3^1\Sigma^+$ and $4^1\Sigma^+$) and ($9^1\Sigma^+$ and $10^1\Sigma^+$). We conclude that the positions of the irregularities in the R-dependence of permanent dipole moments are interrelated to the avoided crossings between the potential energy curves, which are both manifestations of abrupt changes of the character of the electronic wave functions. These crossings have important consequences for the excitation or charge transfer efficiency. For example, neutralization cross sections critically depend on these crossing series and they could be important for charge transfers in various astrophysical conditions [14–17]. Furthermore, we observe that the permanent dipole moment of the $X^1\Sigma^+$, $4^1\Sigma^+$, $5^1\Sigma^+$and $9^1\Sigma^+$states, dissociating into Sr ($5s^2$, 5s4d ,5s5p and 5s6s) + $Na^+$, are significant and show a similar behavior. They yield a nearly linear behavior function of the internuclear distance R, especially for intermediate and large internuclear distances. For the remaining states, dissociating into $Sr^+$ (5s, 5p and 4d) + Na (3s, 3p, 4s and 3d), we remark an important permanent dipole moments in a particular region and then they decrease quickly at large distances. We have also determined the permanent dipole moment for the electronic states of symmetries $^3\Sigma^+$, $^1\Pi$, $^3\Pi$, $^1\Delta$ and $^3\Delta$ and they are shown in figures 8-11. Nothing special for $^3\Sigma^+$ symmetry we observe that the permanent dipole moments of the first ten excited states behave with the same way as in the case of the $^1\Sigma^+$ symmetry. For the remaining $2^3\Sigma^+$, $3^3\Sigma^+$and $6^3\Sigma^+$ electronic states, dissociating respectively into Sr (5*s*5*p*, 5*s*4d and 5*s*6s) + $Na^+$, the permanent dipole moments are significant and product a pure linear behavior function of R. The permanent dipole moments of $8^3\Sigma^+$ and $9^3\Sigma^+$ electronic states, dissociating respectively into $Sr^+$ (5*s*) + Na(3d) and $Sr^+$ (5s) +Na (4*p*), vanishes rapidly at large distance. Similar patterns are observed in the case of the $^1\Pi$, $^3\Pi$, $^1\Delta$ and $^3\Delta$ states. we remark that the permanent dipole moments of these states, one after other, behaves on the same curve and then drops to zero at particular distances corresponding to the avoided crossings between the two neighbor electronic states. To complete this work, the adiabatic transition dipole moments for all possible transitions between the different $^1\Sigma^+$, $^3\Sigma^+$, $^1\Pi$, $^3\Pi$, $^1\Delta$ and $^3\Delta$ molecular states, have been also determined. They are illustrated in figures 12-16. We observe many peaks located at particular distances very close to the avoided crossings position in adiabatic representation. For example in figure 12 , the two threshold respectively for the transition 1-2 and 3-4 for $^1\Sigma^+$ symmetry, around the distances 7 a. u. and 16 a. u. can be related to the change of character of the correspondents states due to the avoided crossing between these states . This function rapidly drops to zero at very large distances. Some transition moments did not drop to zero at large distances but asymptotically reach the corresponding atomic oscillator strength of the allowed atomic transitions taking the example of the transition 5-6 in figure 14. The nature of transition is specified by the selection rules. These accurate transition dipole moments will be used in the near future to evaluate the radiative lifetimes.

## III. Elastic collisions: Results and discussions

### III. 1. Interaction potentials

The potential energy curves of the $1\text{-}3^1\Sigma^+$ electronic states, for the ionic system $(SrNa)^+$ are calculated for a large and dense grid of internuclear distances ranging from 4.5 to 200 a.u. These states dissociate respectively into $Na^+$ + $Sr(5s^2)$, Na(3s) + $Sr^+$(5s), Na(3s) + $Sr^+$(4d). They are displayed in Fig.1. We remark that the ground state has the deepest well compared to the $2^1\Sigma^+$ and $3^1\Sigma^+$ excited states. Their association energies are of the order of several 1000 $cm^{-1}$, which shows the electron delocalization and the formation of a real chemical bound. The spectroscopic properties of these states were discussed in detail in the previous section. Although interaction energy was calculated for a dense grid, analytical form at large and asymptotic limit is essential for the scattering calculations. In the long range where the separation r >$20a_0$ ($a_0$ is the Bohr radius), the potential is given by the sum of the dispersion terms, which in the leading order goes as $1/r^4$. We obtain short-range potentials by pseudopotential method. The short-range part is smoothly combined with the long-range part by spline to obtain the potential for the entire range. We then solve time-independent Schrodinger equation for these potentials with scattering boundary conditions by Numerov-Cooley algorithm. Dissociation energy and equilibrium position are 0.615482 and 6.9 a.u., respectively for $1^1\Sigma^+$ of $(SrNa)^+$. Reduced mass is taken to be 18.211 a.u. The long-range potential is given by the expression:

$$V(r) = -\frac{1}{2}\left(\frac{C_4}{r^4} + \frac{C_6}{r^6} + \ldots\ldots\ldots\right)$$

where $C_4$, $C_6$ correspond to dipole, quadrupole polarizabilities of concerned atom. Hence, the long-range interaction is predominately governed by polarization interaction. Dipole polarizability for $Sr(5s^2)$ is 199 $a_0^3$ and quadrupole polarizability for $Sr(5s^2)$ is 4641 $a_0^5$. By equating the potential to the kinetic energy one can define a characteristic length scale β of the long-range potentials.

$$\beta_4 = \sqrt{\frac{\mu C_4}{\hbar^2}}$$

$\beta_4$ establishes the order of magnitude of the scattering length for the concerned atom-ion potential. In general the scattering length and characteristic length scale for atom-ion system are at least an order of magnitude larger than that of neutral atom-atom collision. The corresponding characteristic energy scale can be written as [18]

$$E^* = \frac{\hbar^2}{2\mu\beta_4^{\,2}}$$

The characteristic energy scale for ion-atom system is at least two orders of magnitude smaller than neutral atom-atom collision. This is one of the reasons behind the challenges on reaching the s-wave scattering regime for ion-atom system than neutral atom-atom case. Characteristic energy scale defines the height of the centrifugal barrier of the effective potential for p-wave ($l = 1$). The position and height of the centrifugal barrier for $l \neq 0$ can be expressed as [18]

$$\beta_{4_l}^{\max} = \sqrt{\frac{2}{l(l+1)}}\beta_4$$

$$E_l^{\max} = \frac{l^2(l+1)^2}{4} E^*$$

So it is evident from the formulae that as $l$ increases the barrier height also increases, which is expected. For $(SrNa)^+$ ground state $1^1\Sigma^+$ the characteristic length is $2570\,a_0$ and characteristic energy scale $E^*$ = 14.96 KHz or 0.0007mK. Now we want to investigate the energy regime for different partial wave collision both theoretically and numerically. We can find the effective potentials for respective $l$. The effective potential for different partial waves ($l$ = 0, 1, 2, ... ) can be expressed as

$$V_{eff}(r) = -\frac{1}{2}\left(\frac{C_4}{r^4} + \frac{C_6}{r^6} + ...\right) + \frac{\hbar^2}{2\mu}\frac{l(l+1)}{r^2}$$

We have plotted the effective potential curves for first five partial waves , i.e. *s*- , *p*- ,*d*-, *f*- and *g*- wave for the ground state $1^1\Sigma^+$ of $(SrNa)^+$ in Fig.17. From the analytical formula above and the numerical effective potential data we have found out the $\beta_{4_l}^{\max}$ and $E_l^{\max}$ values, which are enlisted below in Table 5. We can see that the numerical data's are in excellent agreement with the theoretically predicted ones.

### III. 2 Ion-atom scattering

To investigate the ion-atom scattering we have to obtain the continuum wave function $\psi_l(r)$ of $l^{th}$ partial wave by solving the partial-wave Schrodinger equation.

$$\left[\frac{d^2}{dr^2} + k^2 - \frac{2\mu}{\hbar^2}V(r) - \frac{l(l+1)}{r^2}\right]\psi_l(r) = 0$$

We use Numerov-Cooley algorithm [19] to solve the second- order differential equation as described in previous articles Ref. [20]. We use FORTRAN code to find the wave functions by choosing appropriate boundary condition.

The asymptotic form of $\psi_l(r)$ is

$$\psi_l(r) \approx \sin\left[kr - \frac{l\pi}{2} + \eta_l\right]$$

Here r denotes the ion-atom separation, the wave number k is related to the collision energy E by $E = \frac{\hbar^2 k^2}{2\mu}$ , μ is the reduced mass of the ion-atom pair and $\eta_l$ is the phase shift for $l^{th}$ partial wave.

In Fig. 18 one can notice that the asymptotic form i.e. sinusoidal oscillation of the ground *s*-wave scattering state is obtained at different interparticle separation for different collision energies. At lower energies i.e. at E = 0.1 μK, 1 μK the asymptotic form is achieved at about 50000 $a_0$ and 25000 $a_0$, respectively. At higher collision energy, E = 1 mK the asymptotic form is reached at about 1500 $a_0$ and at energy 1 K the particles shows free oscillation at a quite lower separation, r = 150 $a_0$. For calculating correct phase-shift or scattering length or cross-section data at a particular energy regime one must ensure that they safely reach into the asymptotic regime. The total elastic scattering cross section is expressed as

$$\sigma_{el} = \frac{4\pi}{k^2}\sum_{l=0}^{\infty}(2l+1)\sin^2(\eta_l)$$

The barrier heights are low for the first three, i.e. $l$ = 0, 1, 2 partial waves and hence allow tunneling of the wave function towards the inner region of the barriers. Therefore, we can say that at even low energy, a substantial number of partial waves are going to contribute at total scattering cross-section $\sigma_{el}$ and the number increases as the energy increases. As an example for our system at lower collision energy ~ $\mu$K we need only 2 partial waves for converging $\sigma_{el}$. For energy ~ mK we need as many as 20 partial waves to get a converging result, where for energy ~ 0.1K we need as many as 70 partial waves to reach the convergence for $(SrNa)^+$. So as energy increases more and more number of partial waves start to contribute in the total scattering cross-section, because the centrifugal barrier height corresponding a particular partial wave $l$ is not too high compared to the collision energy and hence the partial waves can tunnel through the barrier and can contribute to the total elastic scattering cross-section. It is worth mentioning here that the centrifugal barrier height for the potential of atom-ion system is quite lower than that of atom-atom system. Thus, large number of partial waves contribute here. In Fig. 19 we plot the partial wave cross-section for $1^1\Sigma^+$ $(SrNa)^+$ against collision energy E (in K). As k →0, the *s*- wave scattering length becomes independent of energy where the *p*- and *d*-wave cross-section varies linearly. According to Wigner threshold laws: as k→0, the phase shift $\eta_l$ for $l^{th}$ partial-wave behaves as $\eta_l \sim k^{2l+1}$ if $l \le (n-3)/2$, with n being the exponent of the long-range potential behaving as $\sim \frac{1}{r^n}$ as $r \to \infty$. If $l > (n-3)/2$ then $\eta_l$ behaves as $\eta_l \sim k^{n-2}$. Since Born-Oppenheimer ion-atom potential goes as $1/r^4$ in the asymptotic limit, as k →0, *s* - wave scattering cross section becomes independent of energy while all other higher partial-wave cross sections go as $\sim k^2$. Fig. 19 demonstrate this fact. We can see the *s* - wave scattering cross-section tends towards a constant value and it is necessary to go into ultra-low regime to show that it is independent of collision energy. Our numerical calculation faces restriction in calculating phase shift data lower than energy $10^{-8}$ K. Hence we choose a different way to valid this point. The near threshold behavior of the elastic scattering phase shift for any partial wave *l* is described by the effective range expansion [21],

$$\lim_{k\to 0} k^{2l+1}\cot\eta_l(k) \sim -\frac{1}{a_l} + \frac{1}{2}r_l k^2 + O(k^4)$$

where $\eta_l(k)$ is the phase shift for *l*-wave at collision energy $E = \frac{\hbar^2 k^2}{2\mu}$. As the contribution of the s-wave becomes dominant towards the threshold, the above expansion is the most useful for *l=0*. For s-wave collision, $a_s$ of the above expression is the scattering length and $r_s$ is the effective range of the potential. In Figure 20, we have plotted $k \cot \eta_0(k)$ vs $k^2$ (in a.u. $k^2 = \frac{E}{2\mu}$) for s-wave collision and we obtain a straight line as expected. Linear fit yields the intercept to be $(1/a_s)$ = 0.00010501, thus $a_s$ = 9522 $a_0$, which implies of having a constant s-wave differential cross-section of 7.9*$10^8$ a.u., which is consistent with our plot in Fig. 19. The *s*-wave partial wave scattering cross-section will become constant to value 8*$10^8$ a.u. (approximately) at further low energy as shown in Fig.19.

As energy increases beyond the Wigner threshold regime, the s-wave scattering length shows a minimum at energy about 10 μK, which may be related to the Ramsauer-Townsend effect as described in ref. [20]. At this minimum, the p- and d- wave elastic scattering cross-sections are finite, hence one can explore p-wave and d-wave interaction at that corresponding energy (μK) regime.

In the Wigner threshold regime, hetero-nuclear ion-atom collisions are dominated by elastic scattering processes since the charge transfer reactions are highly suppressed in this energy regime. Furthermore, resonant charge transfer collisions do not arise in collision of an ion with an atom of different nucleus. In figure 21 we have plotted the total elastic scattering cross-section for the ground state $1^1\Sigma^+$ of $(SrNa)^+$. At high energy limit, the total elastic scattering cross section can be expressed as [22]

$$\sigma_{el} \approx \pi \left( \frac{\mu C_4^2}{\hbar^2} \right)^{1/3} \left( 1 + \frac{\pi^2}{16} \right) E^{-1/3}$$

thus, $\sigma_{el} = cE^{-1/3}$, where c is the proportionality constant expressed as $c = \pi \left( \frac{\mu C_4^2}{\hbar^2} \right)^{1/3} \left( 1 + \frac{\pi^2}{16} \right)$. The logarithm of $\sigma_{el}$ is of the form, $\log_{10} \sigma_{el} = -\frac{1}{3} \log_{10} E + \log_{10} c$, where the slope of the line is -1/3 and $\log_{10} c$ is the intercept which essentially determined by the long range coefficient $C_4$ of the corresponding potential or equivalently by the characteristic length scale β of the potential or by the polarizability of the concerned atom interacting with the ion. The linear fit for $1^1\Sigma^+$ of $(SrNa)^+$ is found to provide the slope -0.331, which is very close to '-1/3' and $\log_{10}$ c = 3.7487 a.u., i.e. c = 5606 a.u. and theoretically calculated c = 5563 a.u. which are in good agreement.

## Conclusion and discussion

In this work, we have carried out a quantum ab initio calculation to illustrate the electronic structure of the ionic molecule $SrNa^+$, in the adiabatic representation. The ab initio approach is based on nonempirical relativistic pseudopotential for the strontium core, complemented by operatorial core valence correlation estimation with parameterized CPP and FCI methods. Extended GTO basis sets have been optimized for both atoms (Sr and Na) to reproduce the experimental energy spectra for 48 atomic levels with a good accuracy. The potential energy curves and their associated spectroscopic constants were computed for the ground and 48 electronic excited states of $^{1,3}\Sigma^+$, $^{1,3}\Pi$, $^{1,3}\Delta$ symmetries. Most of the adiabatic potential energy curves, especially the excited states, are performed for the first time. The spectroscopic constants of the ground and the lower excited states have been compared with the available theoretical [9-10] result. A very good agreement has been observed between our results and those of Aymar et al, but this doesn't prevent to find disagreement concerning the 2 $^3\Delta$ excite state. For a best understanding of the potential energy curves behavior and the electron charge transfer, we have located the avoided crossing positions and we have computed the energy differences at these estimated crossing positions for the $^1\Sigma^+$ symmetry. These avoided crossings are related to a charge transfer between $Sr^+$ and Na, and Sr and $Na^+$ species. To verify the positions of the avoided crossings observed in the potential energy curves, we have calculated and analyzed the spectra of the permanent and transition dipole moments. As it is expected the permanent dipole moments of the electronic states dissociating into Sr ($5s^2$, 5s4d ,5s5p and 5s6s) + $Na^+$ show an almost linear behavior as function of R, especially for intermediate and large internuclear distances. Moreover, the abrupt changes in the permanent dipole moment are localized at particular distances corresponding to the avoided crossings between the two neighbor electronic states. The majority of the reported data for the electronic excited states are performed here for the first time while the good agreement with antecedent calculations for the ground and low lying first excited electronic states gives confidence in the calculation. Our accurate adiabatic potential energy and dipole moment data are made available as supplement material for such investigation and also for further theoretical and experimental uses.

Here we have also studied the elastic scattering between $Sr + Na^+$. We have presented a detailed study of elastic collision over a wide range of energies. Using the accurate potential data for $1^1\Sigma^+$ of $(SrNa)^+$ we first calculate the effective centrifugal potentials. The centrifugal barrier height and position of the highest point of the barrier shows excellent agreement with the theoretically predicted values. Then we have calculated the scattering wave functions with standard Numerov algorithm. The energy normalized scattering wave functions are shown in Fig. 19. The range of asymptotic regime of the ion-atom interaction for different collision energy is clearly shown in that figure. Then we calculate the scattering cross-section for different partial waves (l = 0, 1, 2) for the ground state of our system for a wide range of energies. In the low energy limit ( < 1 mK), elastic scattering cross sections exhibit Wigner law threshold behavior. We have calculated the scattering length for $(SrNa)^+$ ground state and it is 9522 $a_0$. We have also calculated the total elastic scattering cross-section and we find that, in order to get convergent results for the total scattering cross sections for energies of the order ~1 K, one need to take into account at least 100 partial waves. In the high energy limit the total elastic cross section goes as $E^{-1/3}$ and it is verified in the logarithmic plot of $\sigma_{el}$. This low-energy scattering results may be useful for photoassociative formation of cold (SrNa)+ molecular ion. The colliding ground state corresponding to the continuum of $1^1\Sigma^+$ of atom-ion pair $Sr + Na^+$ can be photoassociated to form bound atom-ion pair at 2 $^1\Pi$ state which asymptotically correspond to $Na^+ + Sr(^1D)$ respectively. This transition will only be short range allowed and no triplet state will be accessed in this scheme. At the inner turning point an efficient decay is expected to take place to the low-lying vibrational levels of the ground potential i.e. $1^1\Sigma^+$**.** Hence the possibility of formation of  cold (SrNa)+ can offer new perspective in ultracold chemistry.

**Figure captions**

Figure 1: Adiabatic potential energy curves of the 10 lowest $^1\Sigma^+$ electronic states of $SrNa^+$

Figure 2: Adiabatic potential energy curves of the 10 lowest $^3\Sigma^+$ electronic states of $SrNa^+$

Figure 3: Adiabatic potential energy curves of the 6 lowest $^1\Pi$ (solid line) states of $SrNa^+$

Figure 4: Adiabatic potential energy curves of the 6 lowest $^3\Pi$ (solid line) states of $SrNa^+$

Figure 5: Adiabatic potential energy curves of the three lowest $^1\Delta$ (solid line) states of $SrNa^+$

Figure 6:Adiabatic potential energy curves of the three lowest $^3\Delta$ (solid line) states of $SrNa^+$

Figure 7: Permanent dipole moment of the first ten $^1\Sigma^+$ electronic states of $SrNa^+$

Figure 8: Permanent dipole moment of the first ten $^3\Sigma^+$ electronic states of $SrNa^+$

Figure 9: Permanent dipole moment of the first six $^1\Pi$ electronic states of $SrNa^+$

Figure 10: Permanent dipole moment of the first six $^3\Pi$ electronic states of $SrNa^+$

Figure 11: Permanent dipole moment of the first of $^{1,3}\Delta$ electronic states of $SrNa^+$

Figure 12: Transition dipole moment of the $^1\Sigma^+$ electronic states of $SrNa^+$

Figure 13: Transition dipole moment of the $^3\Sigma^+$ electronic states of $SrNa^+$

Figure 14: Transition dipole moment of the $^1\Pi$ electronic states of $SrNa^+$

Figure 15: Transition dipole moment of the $^3\Pi$ electronic states of $SrNa^+$

Figure 16: Transition dipole moment of the $^{1,3}\Delta$ electronic states of $SrNa^+$

Figure 17: Long range ion-atom potential in miliKelvin (mK) are plotted for s-(black, solid), p-(red dashed), d-(blue dash-dotted), f-(green dash-dot-dotted) and g- (maroon dash-dotted) partial waves for the $1^1\Sigma^+$ state of $SrNa^+$. First five partial wave barriers are shown and the position of the highest points of the corresponding centrifugal barriers is marked as black circles. The values are also listed in Table-5.

Figure 18: Energy-normalized s-wave (l=0) ground scattering wave-functions for $1^1\Sigma^+$ of $SrNa^+$ at different collision energies: (i) when E= 0.1 μK, (ii) E = 1 μK, (iii) E= 1 mK and (iv) for E= 1 K.

Figure 19: s- (solid black), p-(red dashed) and d- (blue dash dotted) wave scattering cross-section in atomic unit (a.u.) are plotted with different collision energy for $1^1\Sigma^+$ of $(SrNa)^+$.

Figure 20: k $\cot\eta_0$ (in atomic unit) for s-wave is plotted with different $k^2$ (atomic unit) values (1 $k^2$ = ) for $1^1\Sigma^+$ state of $(SrNa)^+$.

Figure 21: Logarithm of total scattering cross section in atomic unit are plotted as a function of logarithm of energy E in atomic unit for $1^1\Sigma^+$ state of $(SrNa)^+$.

**Table 1**. l-dependant cut-off parameter for Strontium and sodium atoms (in a.u.).

| *l* | Sr | [10] | Na | [10] |
|---|---|---|---|---|
| *s* | 2.082 | 2.130 | 1.442 | 1.450 |
| *p* | 1.919 | 2.183 | 1.625 | 1.645 |
| *d* | 1.644 | 1.706 | 1.500 | 1.500 |

**Table 2:** Asymptotic energies of the $SrNa^+$ electronic states: comparison between our energy and the experimental [23] and theoretical [9-10] energies. $\Delta E_1$ is the difference with the experimental dissociation limits and $\Delta E_2$ is the difference with the theoretical dissociation limits.

| State | Our work(a.u.) | [23](a. u.) | [10](a.u.) | $\Delta E_1(cm^{-1})$ | $\Delta E_2(cm^{-1})$ |
|---|---|---|---|---|---|
| $X^1\Sigma^+$ | -0.615027 | -0.61464148 | -0.615471 | 84.71 | 97.44 |
| $2^1\Sigma^+$ | -0.594202 | -0.594215 | -0.594215 | 2.85 | 2.85 |
| $3^1\Sigma^+$ | -0.527247 | -0.52283208 | -0.520684 | 968.98 | 1440.41 |
| $4^1\Sigma^+$ | -0.520527 | -0.51577530 | -0.517756 | 1042.94 | 608.16 |
| $5^1\Sigma^+$ | -0.517754 | -0.517756 | | 0.43 | |
| $6^1\Sigma^+$ | -0.516898 | -0.516904 | | 1.31 | |
| $7^1\Sigma^+$ | -0.484326 | -0.483723 | | 132.34 | |
| $8^1\Sigma^+$ | -0.476920 | -0.476945 | | 5.48 | |
| $9^1\Sigma^+$ | -0.475244 | -0.475073 | | 37.53 | |
| $10^1\Sigma^+$ | -0.461282 | -0.461289 | | 1.53 | |
| $1^3\Sigma^+$ | -0.594202 | -0.594215 | -0.594215 | 2.85 | 2.85 |
| $2^3\Sigma^+$ | -0.548331 | -0.54765155 | -0.548880 | 149.24 | 120.5 |
| $3^3\Sigma^+$ | -0.529928 | -0.53147099 | -0.529868 | 338.43 | 13.16 |
| $4^3\Sigma^+$ | -0.527247 | -0.527120 | -0.527120 | 27.87 | 27.87 |
| $5^3\Sigma^+$ | -0.516898 | -0.516904 | | 1.31 | |
| $6^3\Sigma^+$ | -0.481060 | -0.482320 | | 276.53 | |
| $7^3\Sigma^+$ | -0.484326 | -0.483723 | | 132.34 | |
| $8^3\Sigma^+$ | -0.476920 | -0.476945 | | 5.48 | |
| $9^3\Sigma^+$ | -0.461282 | -0.461289 | | 1.53 | |
| $10^3\Sigma^+$ | -0.456275 | -0.456284 | | 1.97 | |
| $1^1\Pi$ | -0.527247 | -0.52283208 | -0.520684 | 968.98 | 1440.41 |
| $2^1\Pi$ | -0.520527 | -0.51577530 | -0.517756 | 1042.94 | 608.16 |
| $3^1\Pi$ | -0.517754 | -0.517756 | | 0.43 | |
| $4^1\Pi$ | -0.516898 | -0.516904 | | 1.31 | |
| $5^1\Pi$ | -0.484326 | -0.483723 | | 132.34 | |
| $6^1\Pi$ | -0.461282 | -0.461289 | | 1.53 | |
| $1^3\Pi$ | -0.548331 | -0.54765155 | -0.548880 | 149.24 | 120.5 |
| $2^3\Pi$ | -0.529928 | -0.53147099 | -0.529868 | 338.43 | 13.16 |
| $3^3\Pi$ | -0.527247 | -0.527120 | -0.527120 | 27.87 | 27.87 |
| $4^3\Pi$ | -0.516898 | -0.516904 | | 1.31 | |
| $5^3\Pi$ | -0.484326 | -0.483723 | | 132.34 | |
| $6^3\Pi$ | -0.461282 | -0.461289 | | 1.53 | |
| $1^1\Delta$ | -0.527247 | -0.52283208 | -0.520684 | 968.98 | 1440.41 |
| $2^1\Delta$ | -0.520527 | -0.518634 | -0.520730 | 415.46 | 44.55 |
| $3^1\Delta$ | -0.461282 | -0.461289 | | 1.53 | |
| $1^3\Delta$ | -0.529928 | -0.53147099 | -0.529868 | 347 | 3.95 |
| $2^3\Delta$ | -0.527247 | -0.527120 | -0.527120 | 338.43 | 13.16 |
| $3^3\Delta$ | -0.461282 | -0.461289 | | 1.53 | |

**Table 3:** Spectroscopic constants for the first eight electronic states of $SrNa^+$ ionic molecule

| State | $R_e$(a. u.) | $D_e$($cm^{-1}$) | $T_e$($cm^{-1}$) | $\omega_e$($cm^{-1}$) | $\omega_e x_e$($cm^{-1}$) | $B_e$($cm^{-1}$) | Ref. |
|---|---|---|---|---|---|---|---|
| $X^1\Sigma^+$ | 6.92 | 8408 | 0 | 118.02 | 0.37 | 0.068970 | This work |
| | 6.9 | 8499 | | 121 | | | [9] |
| | 6.85 | 8361.57 | | 120.82 | | | [11] |
| $2^1\Sigma^+$ | 14.85 | 323 | 12645 | 21.06 | 0.39 | 0.014983 | This work |
| | 14.8 | 359 | | 21 | | | [9] |
| | 14.73 | 327.41 | 19546.21 | 20.98 | | | [11] |
| $1^3\Sigma^+$ | 8.03 | 4422 | 8544 | 83.07 | 0.32 | 0.051244 | This work |
| | 8.03 | 4465 | | 83 | | | [9] |
| | 7.871 | 4549.93 | 9042 | 16 | | | [11] |
| | 8.030 | 4465.00 | | | | | [11] |
| $2^3\Sigma^+$ | 7.40 | 2863 | 20053 | 96.46 | 0.38 | 0.060401 | This work |
| | 7.4 | | | 98 | | | [9] |
| | 7.28 | 3915.71 | 18 960.22 | 102.86 | | | [11] |
| | 13.06 | 2367 | 20549 | 39.39 | 0.38 | 0.060401 | This work |
| | 13.0 | | | 40 | | | [9] |
| | 13.24 | 2102.53 | | | | | [11] |
| | 13.00 | –2428.00 | | | | | [11] |
| $1^1\Pi$ | 7.64 | 5850 | 21810 | 88.79 | 0.29 | 0.056595 | This work |
| | 7.65 | 5926 | | 90 | | | [9] |
| | 7.51 | 6450.43 | 21438.08 | 87.45 | | | [11] |
| | 7.16 | 3249.15 | 26176.54 | 96.35 | | | [11] |
| $1^3\Pi$ | 6.54 | 7745 | 15187 | 129.05 | 0.67 | 0.077274 | This work |
| | 6.58 | 7677 | | 127 | | | [9] |
| | 6.36 | 7617.96 | 15 632.98 | 133.97 | | | [11] |
| | 7.66 | 5717.68 | 21499.54 | 84.89 | | | [11] |
| $1^1\Delta$ | 7.38 | 7431 | 20231 | 98.33 | 0.32 | 0.060634 | This work |
| | 7.4 | 7545 | | 100 | | | [9] |
| | 7.35 | 7665.65 | 20145.38 | 94.60 | | | [11] |
| $1^3\Delta$ | 7.42 | 7154 | 19933 | 97.44 | 0.33 | 0.060068 | This work |
| | 8.9 | 7545 | | 72 | | | [9] |
| | 7.40 | 7049.56 | 19971.99 | 95.00 | | | [11] |

**Table 4:** Spectroscopic constants of the excited singlet and triplet electronic states of $SrNa^+$

| State | $R_e$(a. u.) | $D_e$($cm^{-1}$) | $T_e$($cm^{-1}$) | $\omega_e$($cm^{-1}$) | $\omega_e x_e$($cm^{-1}$) | $B_e$($cm^{-1}$) | Ref. |
|---|---|---|---|---|---|---|---|
| $3^1\Sigma^+$ | | | | | | | |
| First min | 7.93 | 5333 | 22330 | 84.19 | 0.23 | 0.052598 | |
| | 7.9 | 5488 | | 85 | | | [9] |
| | 8.18 | 4992.25 | 22331.35 | 100.43 | | | [11] |
| | 14.6 | 834.55 | | 84.46 | | | [11] |
| $4^1\Sigma^+$ | | | | | | | |
| First min | 8.07 | 2243 | 27018 | 124.52 | 0.97 | 0.050818 | |
| | 8.15 | 2068 | | 128 | | | [9] |
| | 8.08 | 253300 | 28790.48 | 31.89 | | | [11] |
| Second min | 13.40 | 2572 | 26688 | 49.73 | 0.97 | 0.050818 | |
| | 13.6 | 2472 | | | | | [9] |
| | 14.44 | 2491.31 | | | | | [11] |
| $5^1\Sigma^+$ | | | | | | | |
| First min | 9.93 | 867 | 28873 | 94.11 | 1.09 | 0.033526 | |
| | 9.88 | 1119.94 | 32741.03 | 146.55 | | | [11] |
| Second min | 16.16 | 497 | 29243 | 13.35 | 1.09 | 0.033526 | |
| $6^1\Sigma^+$ | | | | | | | |
| First hump | 6.51 | -10490 | | | | | |
| | 6.67 | –10089.21 | 40640 | 5 9.5 | | | [11] |
| | 24.52 | 400.10 | 40640.5 | | | | [11] |
| $7^1\Sigma^+$ | | | | | | | |
| First min | 13.15 | 569 | 36514 | 41.85 | 0.94 | 0.019109 | |

| | | | | | | |
|---|---|---|---|---|---|---|
| | 13.55 | 443.65 | 42608.61 | 31.81 | | [11] |
| Second min | 23.29 | 337 | 36746 | 19.36 | 0.94 | 0.019109 |
| $8^1\Sigma^+$ | | | | | | |
| First min | 21.18 | 1699 | 37006 | 15.75 | 5.90 | 0.007342 |
| | 7.38 | 3652.3 | 43293.37 | 20.74 | | [11] |
| | 22.86 | 1633.48 | | | | [11] |
| $9^1\Sigma^+$ | | | | | | |
| First hump | 9.98 | -2475 | | | | |
| | 9.14 | –2911.80 | 42957 | 42 4.9 | | [11] |
| | 14.75 | –1499.98 | | | | [11] |
| First min | 35.51 | 255 | 38876 | 7.71 | 1.19 | 0.033171 |
| | 35.60 | 57.62 | | | | [11] |
| $10^1\Sigma^+$ | | | | | | |
| First hump | 9.78 | -778 | | | | |
| | 11.64 | –938.9 43 | 361.69 | 116.82 | | [11] |
| | 19.54 | 1340.19 | | | | [11] |
| Second hump | 16.96 | -330 | | | | |
| First min | 31.30 | 910 | 41211 | 9.03 | 1.10 | 0.034587 |
| | 28.14 | 102.19 | | | | [11] |
| $2^1\Pi$ | | | | | | |
| First min | 7.32 | 3378 | 25875 | 89.97 | 0.77 | 0.061780 |
| | 7.3 | 3286 | | 90 | | [9-10] |
| | 9.37 | 755.23 | 31661.21 | 63.65 | | [11] |
| $3^1\Pi$ | | | | | | |
| First min | 9.44 | 350 | 29401 | 53.14 | 1.03 | 0.037108 |
| | | | 39296.73 | | | [11] |
| Second min | 14.71 | 246 | 29505 | 18.42 | 1.03 | 0.037108 |
| $4^1\Pi$ | | | | | | |
| | | | Repulsive | | | |
| | 13.52 | 546.55 | 41509.04 | 29.50 | | [11] |
| $5^1\Pi$ | | | | | | |
| First min | 13.32 | 589 | 36491 | 31.10 | 0.47 | 0.018645 |
| | 9.41 | 1139.60 | 43159.49 | 45.02 | | [11] |
| | 17.71 | 1192.59 | | | | [11] |
| $6^1\Pi$ | | | | | | |
| First min | 9.70 | 890 | 41126 | 50.79 | 2.27 | 0.035095 |
| | 10.51 | 965.91 | 44366.60 | 35.42 | | [11] |
| Second min | 19.43 | 1523 | 40494 | 19.48 | 2.27 | 0.035095 |
| | 16.76 | 966.50 | | | | [9] |
| $2^1\Delta$ | | | | | | |
| First min | 16.19 | 220 | 29043 | 17.91 | 0.38 | 0.012605 |
| | 15.64 | 211.29 | 37957.94 | 7.945 | | [11] |
| $3^1\Delta$ | | | | | | |
| First min | 8.17 | 527 | 41423 | 61.46 | 2.08 | 0.049554 |
| | 10.03 | 345.29 | 42180.63 | 97.38 | | [11] |
| Second min | 14.39 | 1548 | 40402 | 30.09 | 2.08 | 0.049554 |
| | 28.30 | 202.70 | | | | [11] |
| $3^3\Sigma^+$ | | | | | | |
| First min | 9.90 | 4317 | 22766 | 150.07 | 2.56 | 0.033706 |
| | 9.9 | 4303 | | 146 | | [9] |
| | 10.12 | 4397.20 | 22331.35 | 144.43 | | [11] |
| $4^3\Sigma^+$ | | | | | | |
| First min | 17.82 | 160 | 27502 | 13.28 | 0.38 | 0.010410 |
| | 17.4 | 160 | | 42 | | [9] |
| | 17.36 | 338.45 | 34964.31 | 15.68 | | [11] |
| $5^3\Sigma^+$ | | | | | | |
| First min | 18.02 | 151 | 27511 | 13.07 | 0.28 | 0.010180 |
| | 16.66 | 713.22 | 42296.95 | 38.04 | | [11] |
| $6^3\Sigma^+$ | | | | | | |
| First min | 18.94 | 826 | 29093 | 20.67 | 0.12 | 0.009216 |
| | 12.84 | 507.32 | 42841.25 | 22.73 | | [11] |
| | 21.86 | 630.66 | | | | [11] |
| $7^3\Sigma^+$ | | | | | | |
| First min | 12.44 | 644 | 36438 | 30.94 | 0.39 | 0.021355 |
| | 8.7 | –2323.73 | 42 948.79 | 13.87 | | [11] |
| Second min | 22.49 | 1149 | 35934 | 20.36 | 0.39 | 0.021355 |
| | 19.29 | 1976.38 | | | | [11] |
| $8^3\Sigma^+$ | | | | | | |
| First min | 18.57 | 611 | 36951 | 21.38 | 0.82 | 0.009582 |
| | 10.71 | –594.244 | 998.69 | 16.05 | | [11] |
| | 27.63 | 856.38 | | | | [11] |
| $9^3\Sigma^+$ | | | | | | |
| First min | 13.84 | -1638 | 40326 | 32.51 | 0.07 | 0.017262 |
| | 10.9 | –300.45 | 46019.24 | 11.73 | | [11] |
| | 19.46 | 882.98 | | | | [11] |
| Second min | 31.76 | 339 | 38349 | 9.85 | 0.07 | 0.017262 |
| | 30.84 | 857.43 | | | | [11] |
| $10^3\Sigma^+$ | | | | | | |
| First min | 14.53 | 1474 | 40482 | 29.34 | 0.36 | 0.015661 |
| | 12.53 | 154.28 | 47101.25 | 76.58 | | [11] |
| | 15.65 | 537.99 | | | | [11] |
| Second min | 30.12 | 976 | 40980 | 10.99 | 0.36 | 0.015661 |
| | 22 | 3611.67 | | | | [11] |
| $2^3\Pi$ | | | | | | |
| First min | 7.79 | 5586 | 21501 | 85.12 | 0.42 | 0.054448 |
| | 7.8 | 5631 | | 86 | | [9] |
| | 17.21 | 180.96 33 | 996.42 | 12.92 | | [11] |
| $3^3\Pi$ | | | | | | |
| First min | 17.88 | 152 | 27511 | 13.17 | 0.38 | 0.010340 |
| | 17.7 | 135 | | 13 | | [9] |
| | | | 38104.99 | | | [11] |
| $4^3\Pi$ | | | | | | |
| | | | Repulsive | | | |
| | 10.34 | 1740.72 | 40251.45 | 53.46 | | [11] |
| $5^3\Pi$ | | | | | | |
| First min | 10.42 | 1922 | 35161 | 59.16 | 0.44 | 0.030462 |
| | 8.91 | 1227.02 | 42411.08 | 30.56 | | [11] |
| | 14.15 | 1856.20 | | | | [11] |
| $6^3\Pi$ | | | | | | |
| First min | 8.32 | 1565 | 40392 | 49.37 | 0.32 | 0.047714 |
| | 9.88 | 1225.52 | 42665.67 | 12.46 | | [11] |
| Second min | 13.72 | 1693 | 40264 | 28.34 | 0.32 | 0.047714 |
| Third min | 19.51 | 1505 | 40452 | 20.54 | 0.32 | 0.047714 |
| | 18.96 | 1356.37 | | | | [11] |
| $2^3\Delta$ | | | | | | |
| First min | 18.02 | 152 | 27511 | 13.09 | 0.32 | 0.010177 |
| | 17.31 | 152.04 | 20145.38 | 94.60 | | [11] |
| $3^3\Delta$ | | | | | | |
| First min | 13.84 | 1631 | 40326 | 32.63 | 0.06 | 0.017258 |
| | 7.93 | 694.15 | 41781.19 | 31.52 | | [11] |
| | 14.56 | 1835.14 | | | | [11] |

**Table 5:** Theoretical and numerical value (from numerical effective potential data) of height and position of the highest point of the barrier for different partial waves

| Partial wave $l$ | Theoretical $\beta_{4_l}^{\max}$ ($a_0$) | Numerical $\beta_{4_l}^{\max}$ ($a_0$) | Theoretical $E_l^{\max}$ (mK) | Numerical $E_l^{\max}$ (mK) |
|---|---|---|---|---|
| 1 | 2570.24 | 2570.25 | 0.0007 | 0.0007 |
| 2 | 1483.93 | 1483.94 | 0.0063 | 0.0065 |
| 3 | 1049.30 | 1049.31 | 0.0252 | 0.0259 |
| 4 | 812.78 | 812.80 | 0.07 | 0.0720 |

**Figure 1**

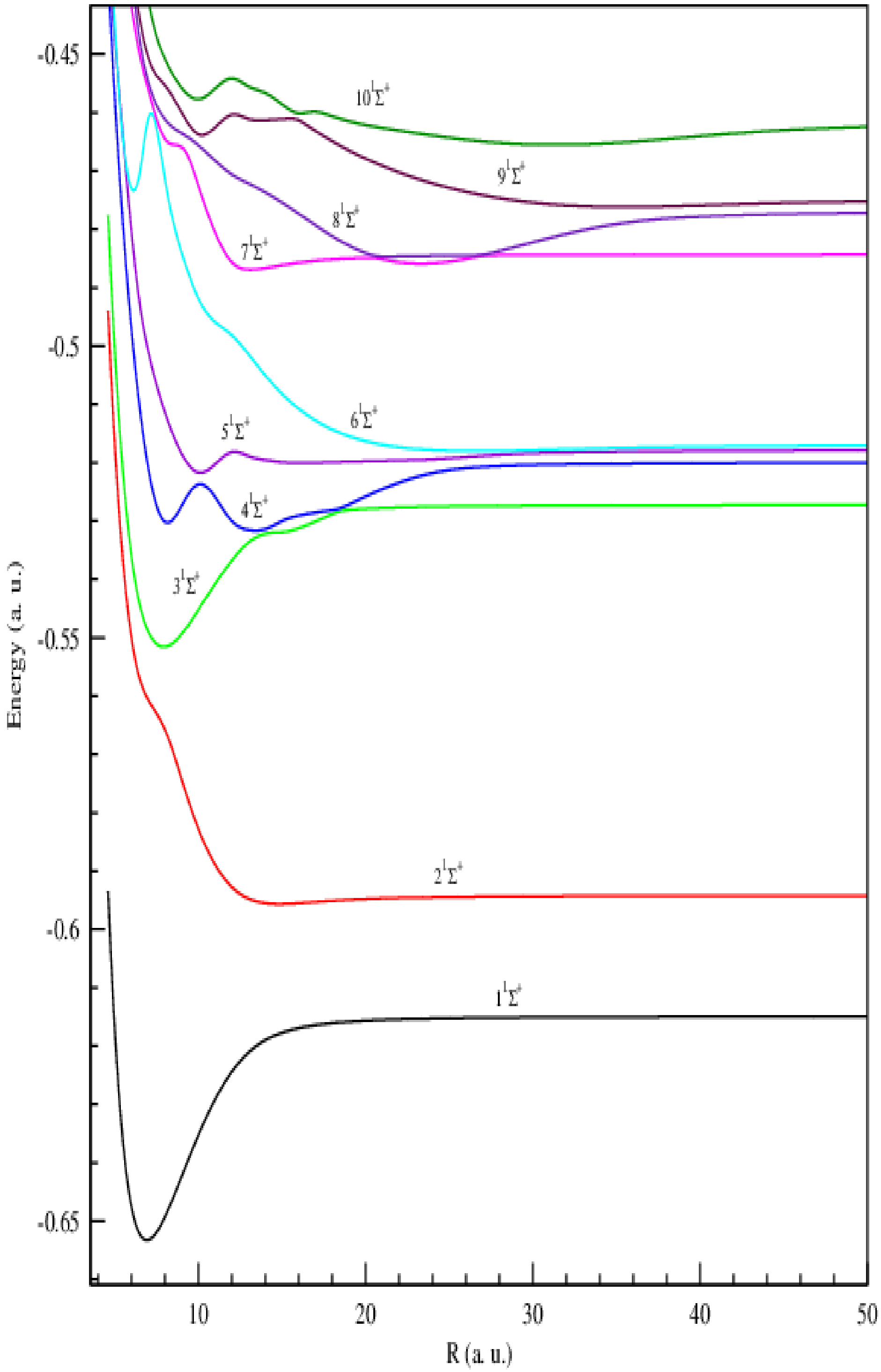

**Figure 2**

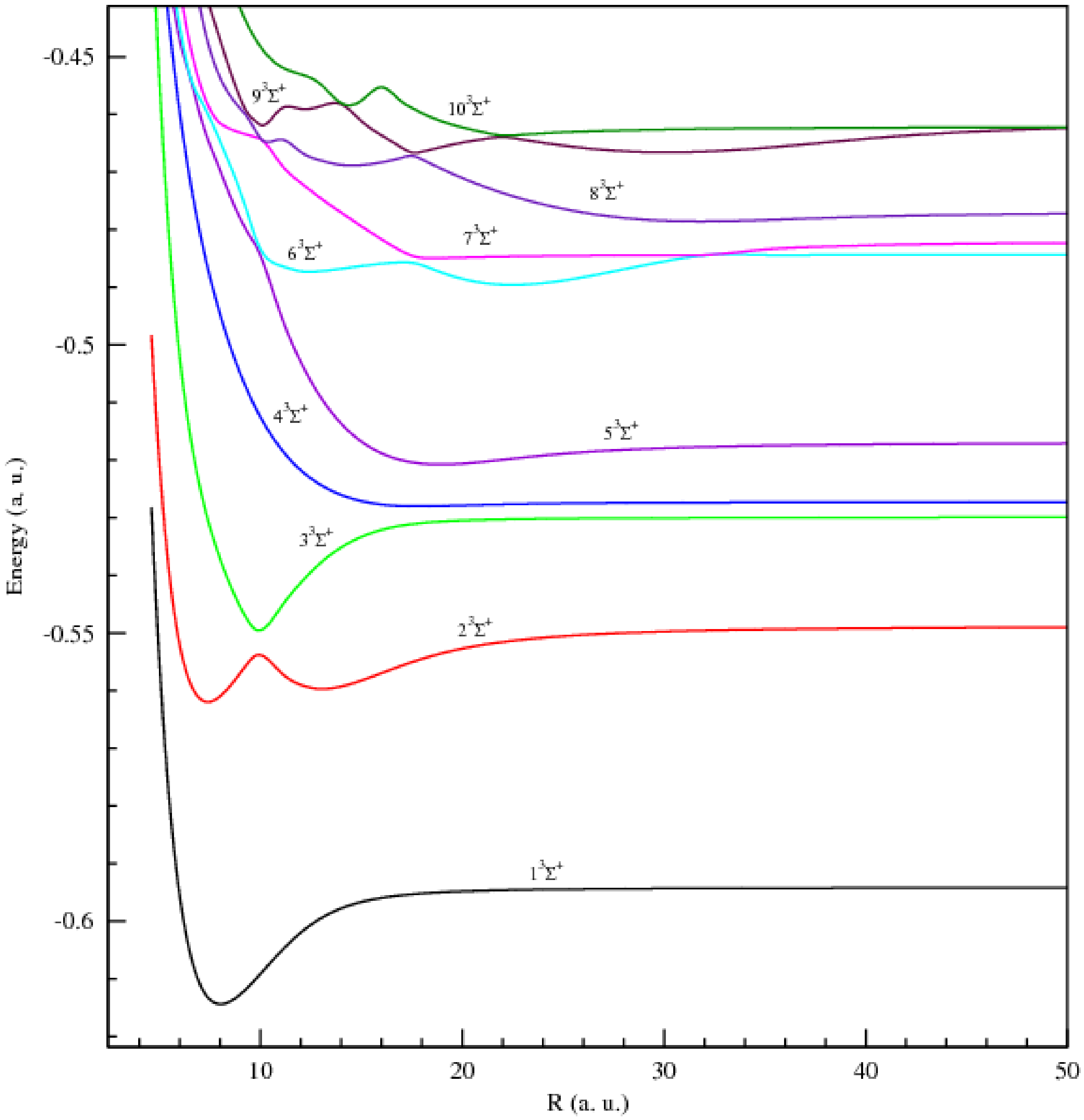

**Figure 3:**

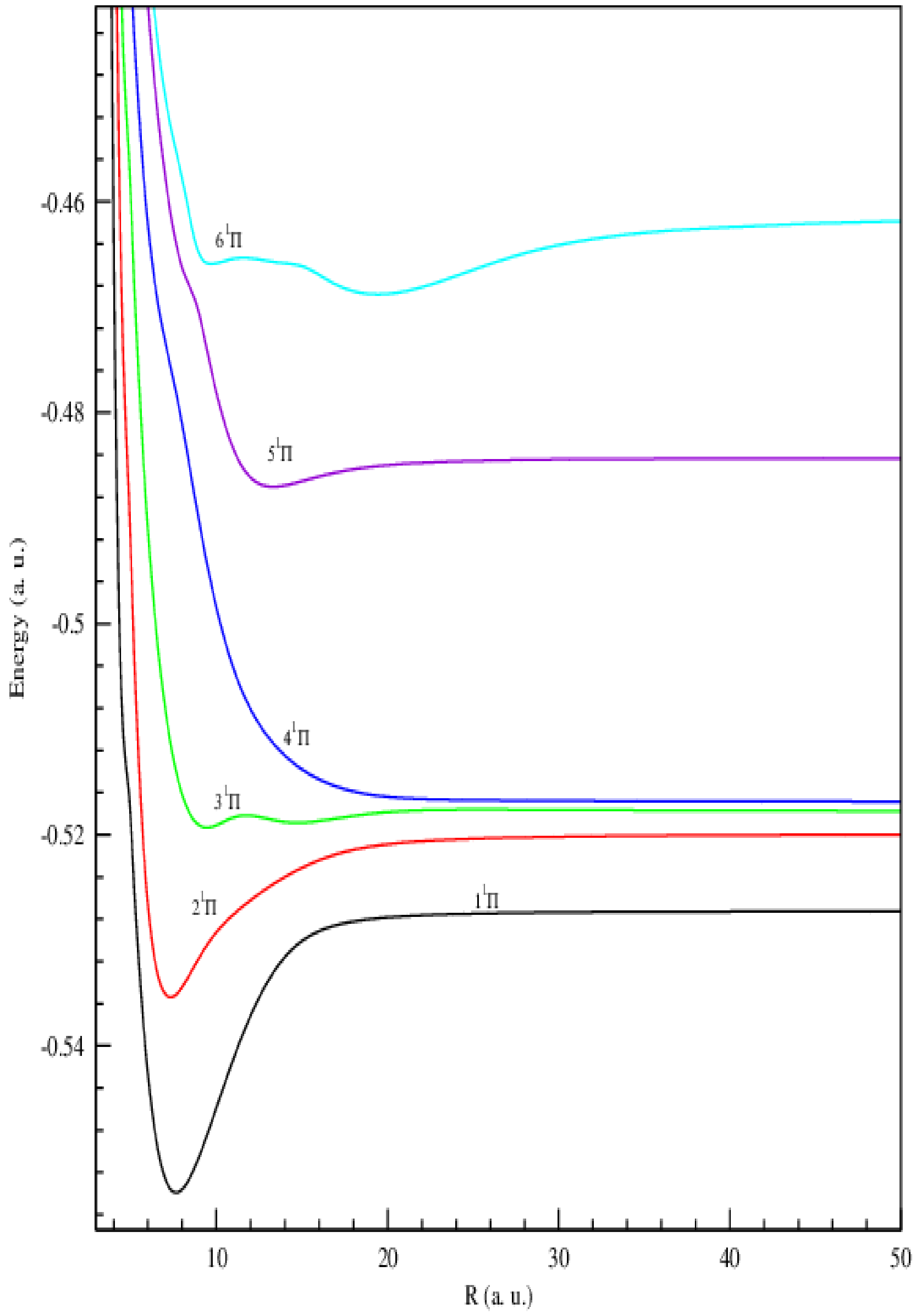

**Figure 4**

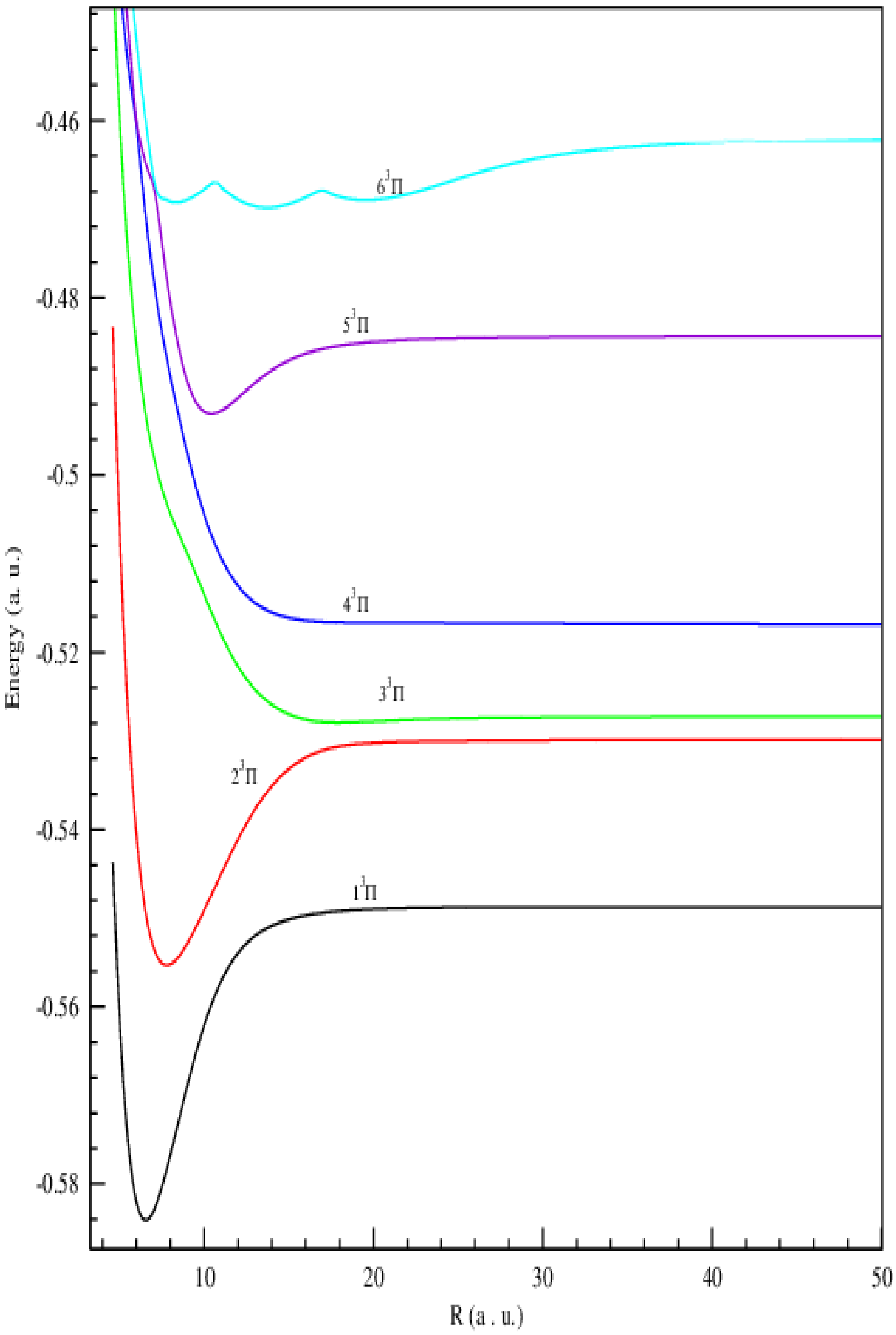

**Figure 5**

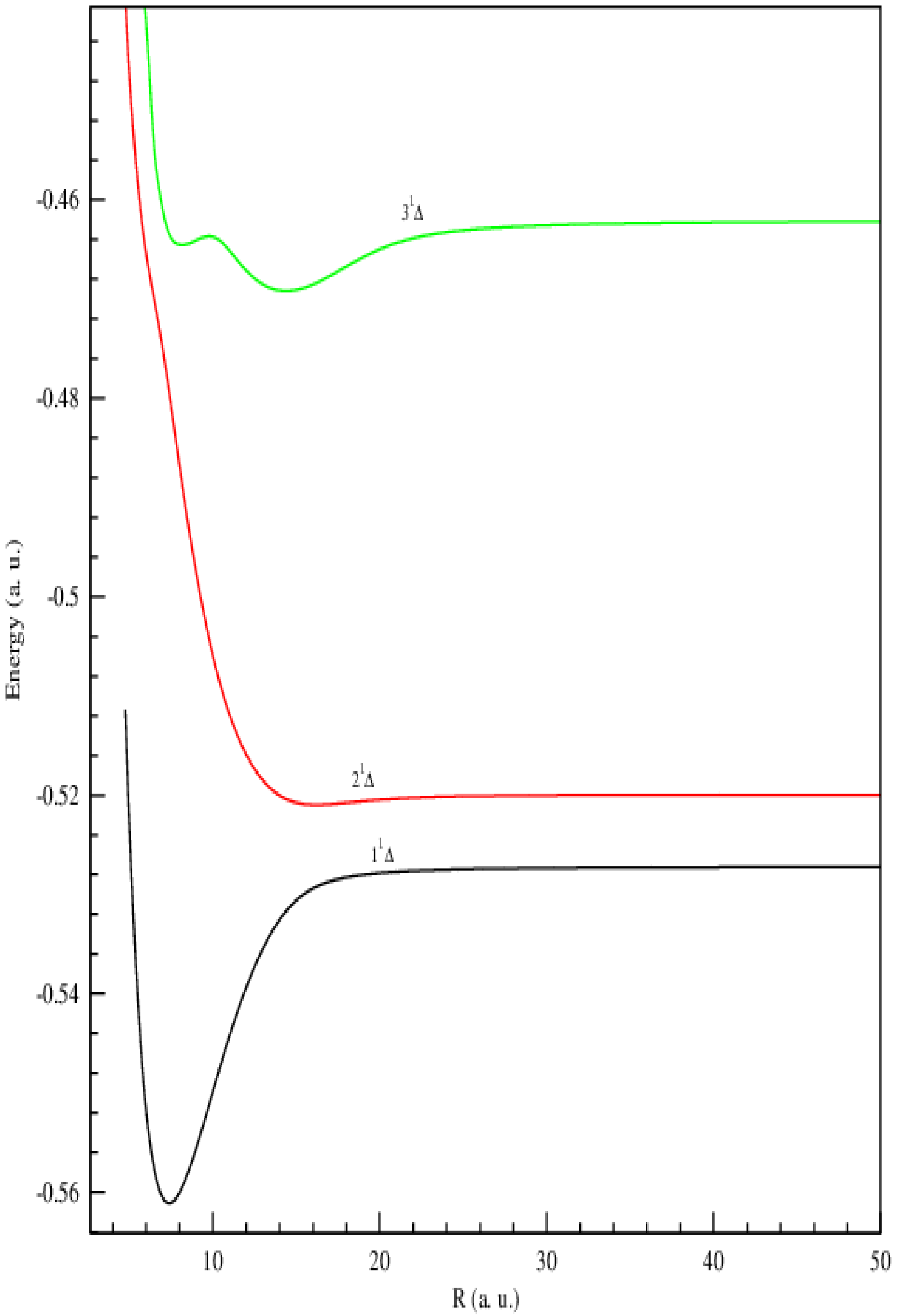

**Figure 6**

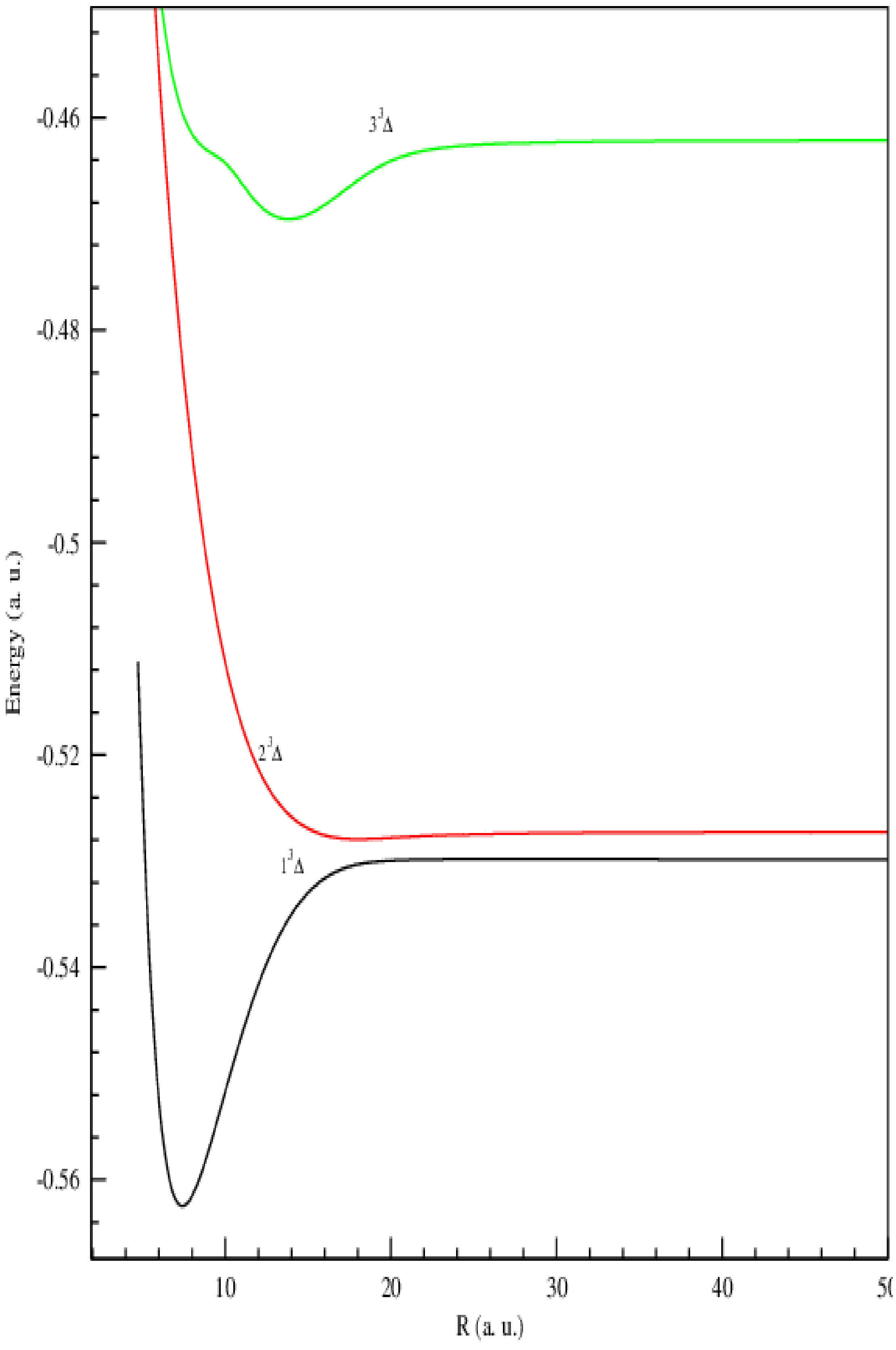

# Figure 7

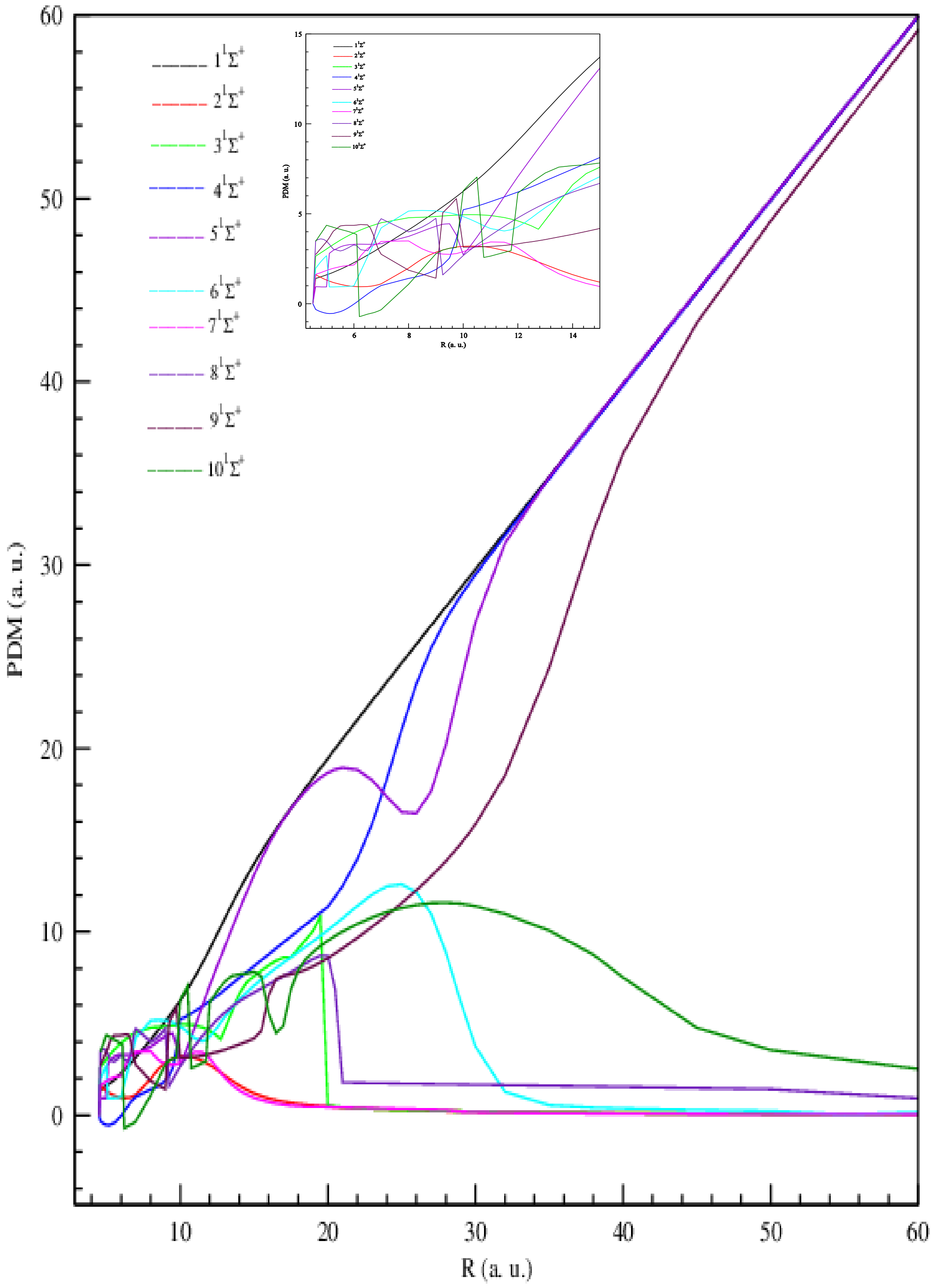

**Figure 8**

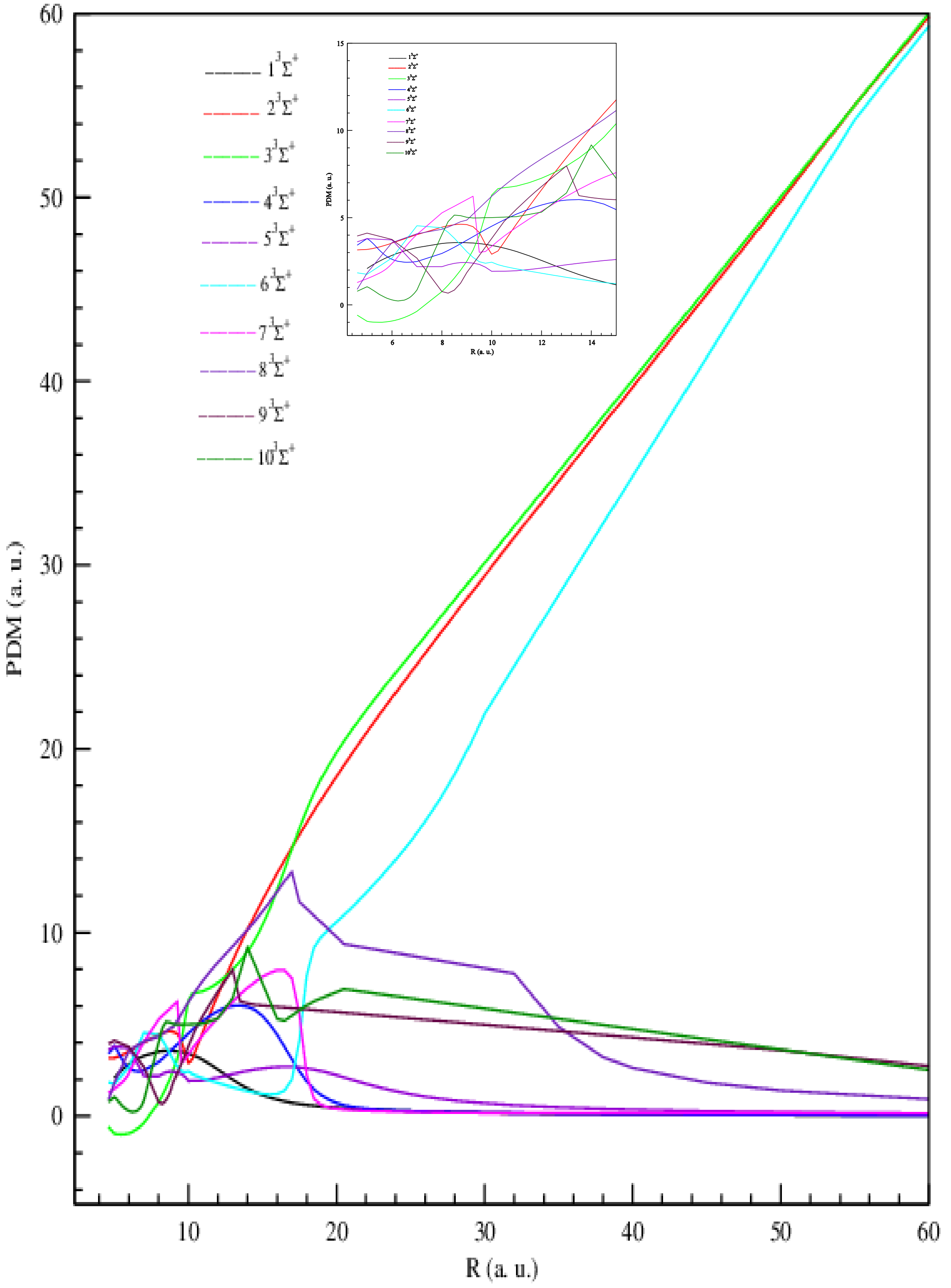

**Figure 9**

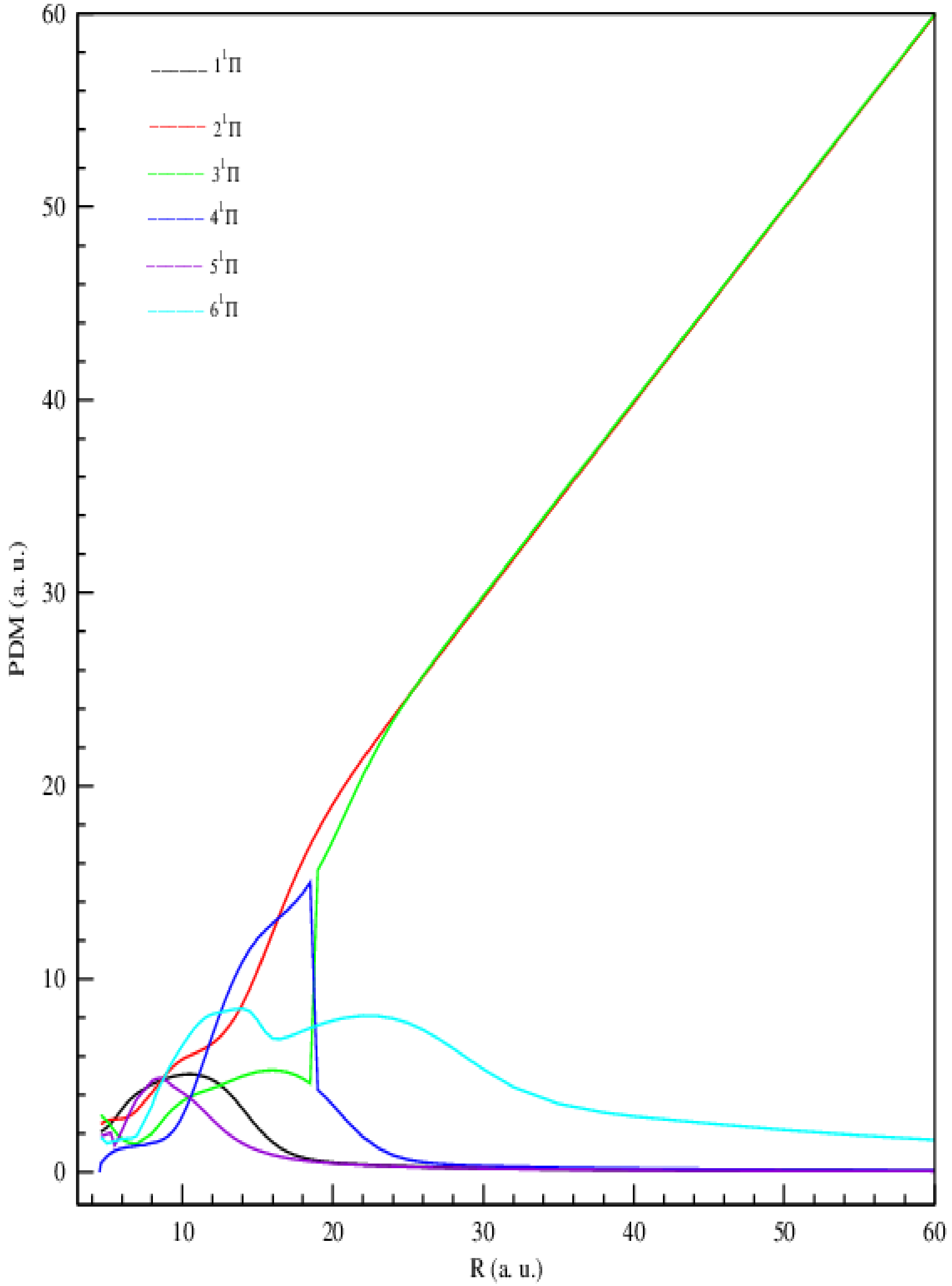

**Figure 10**

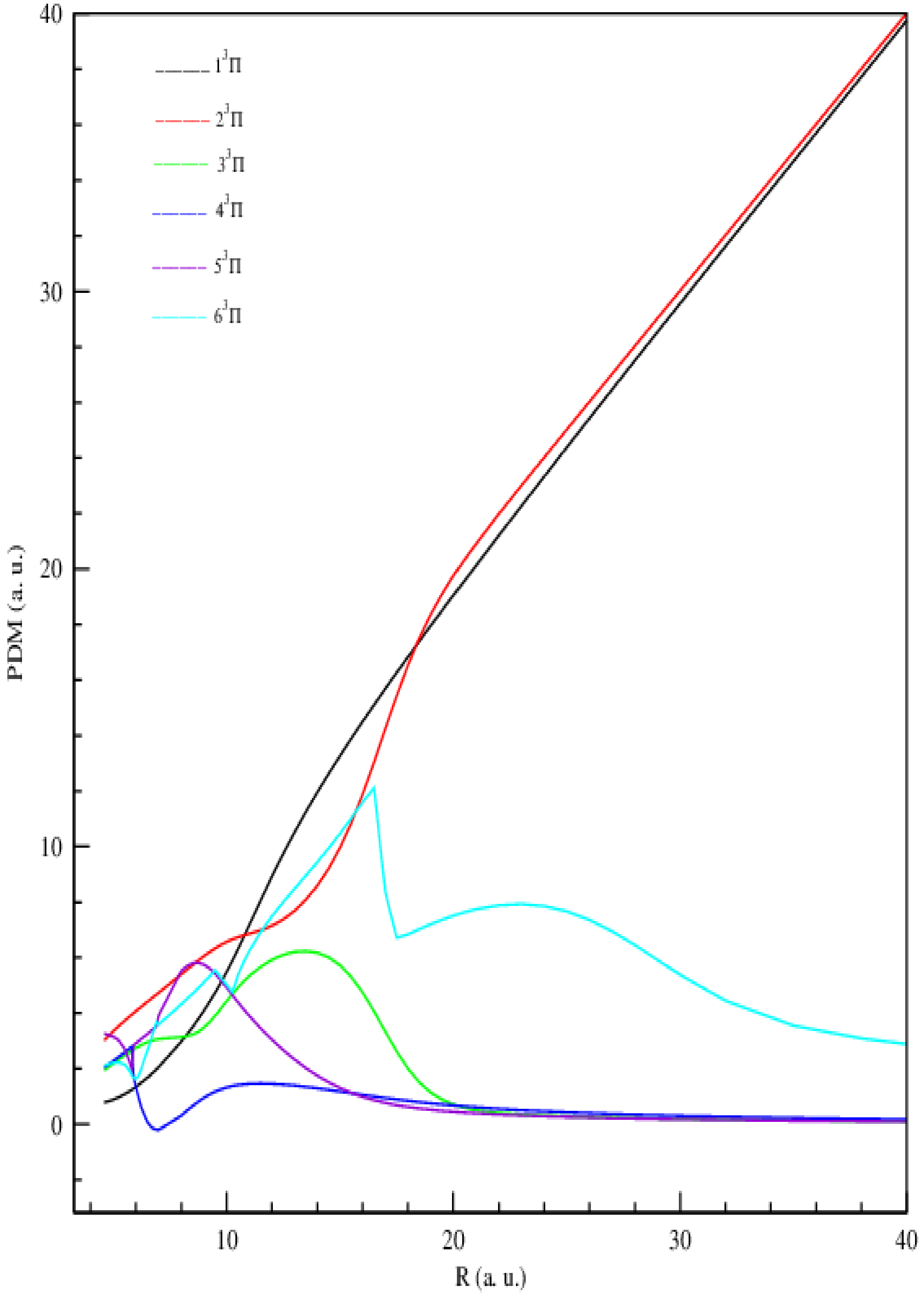

**Figure 11**

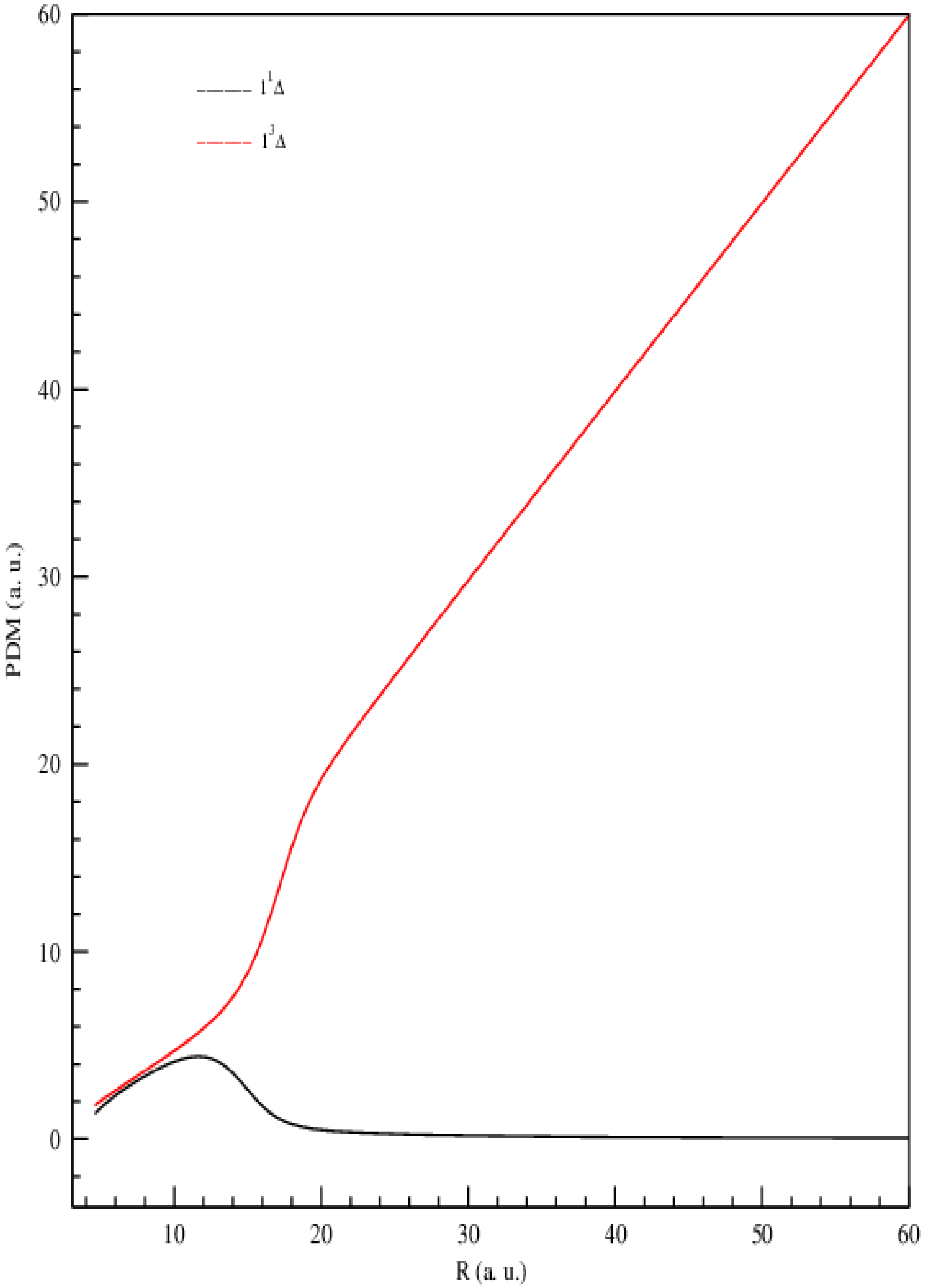

**Figure 12**

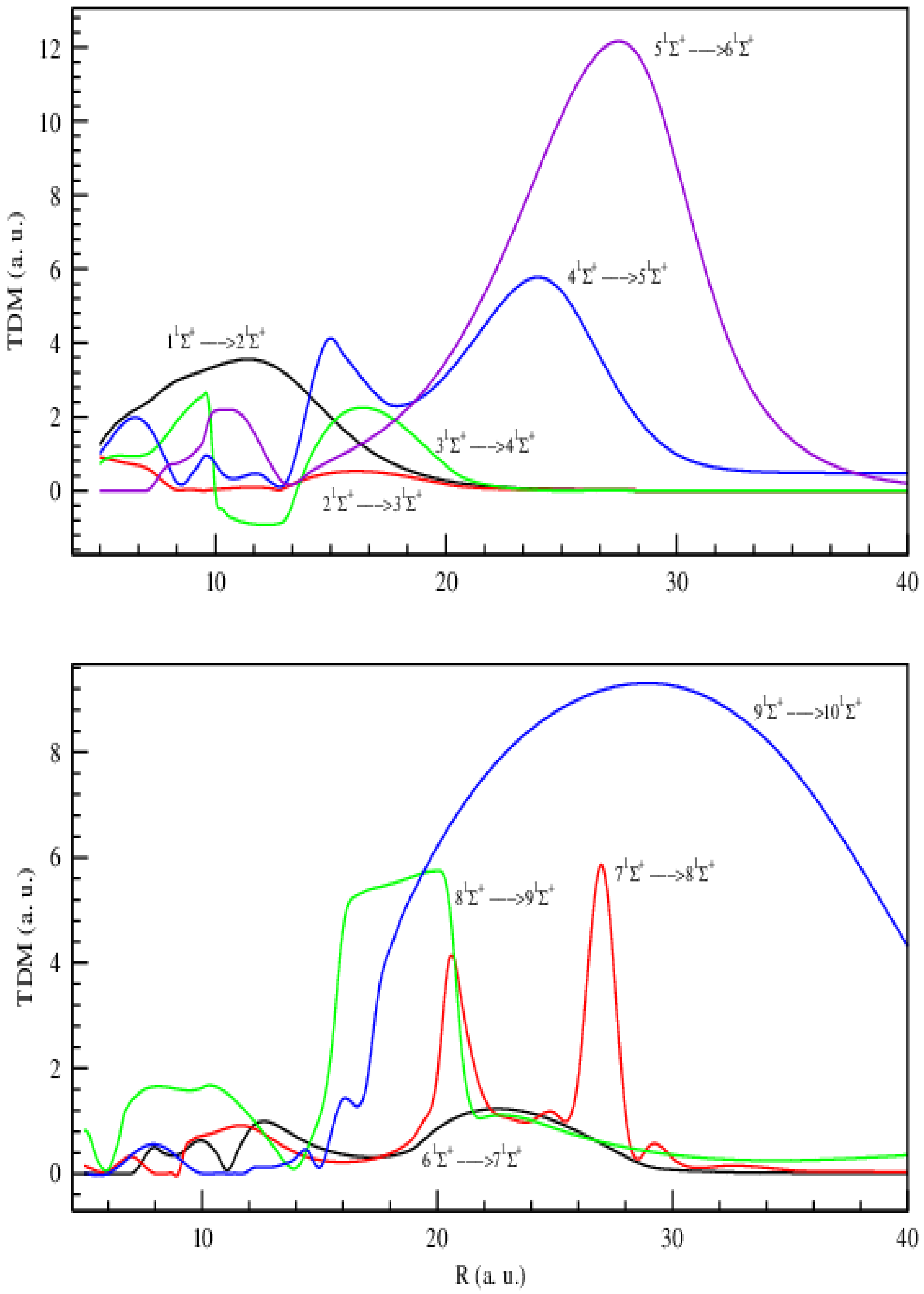

# Figure 13

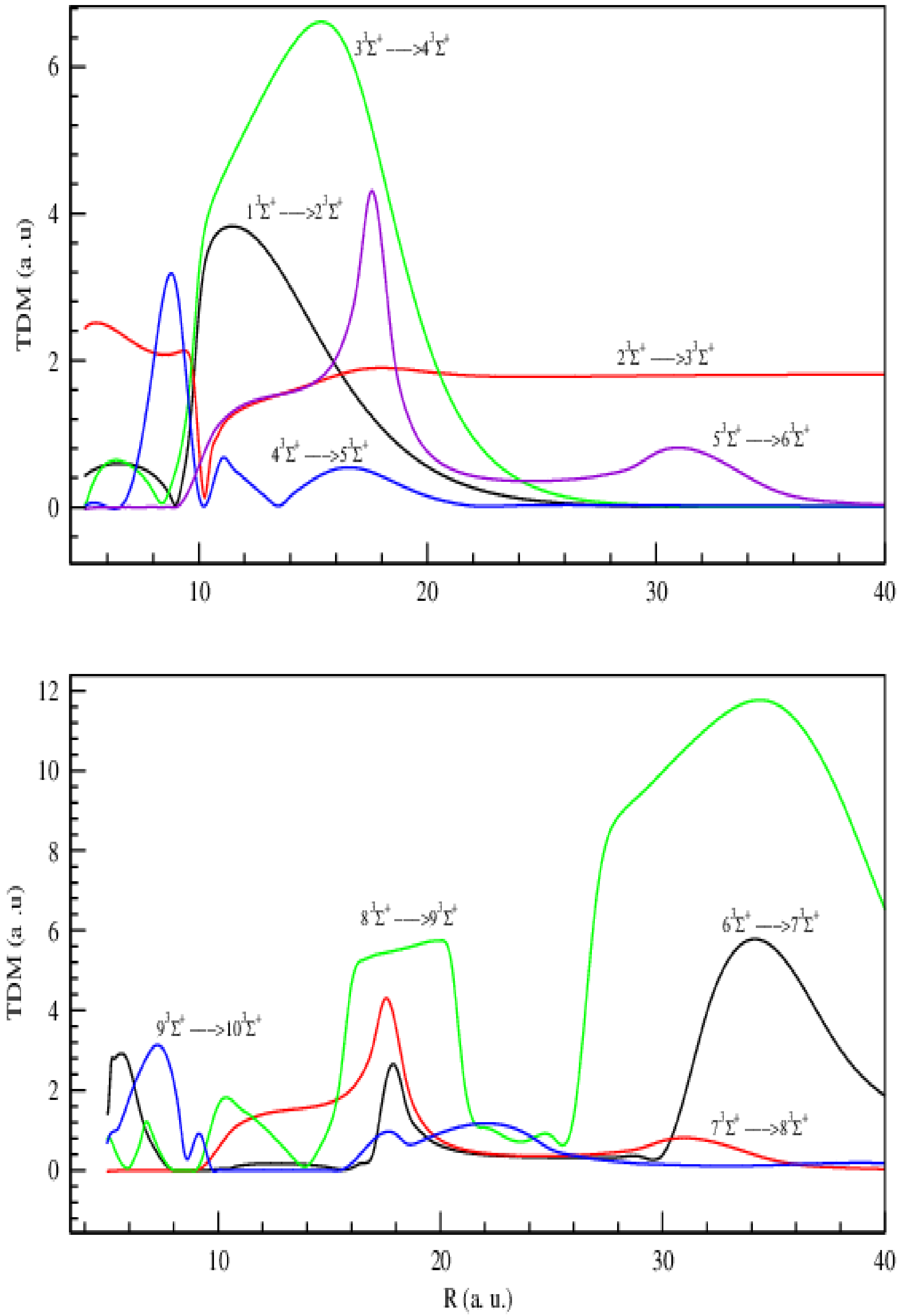

**Figure 14**

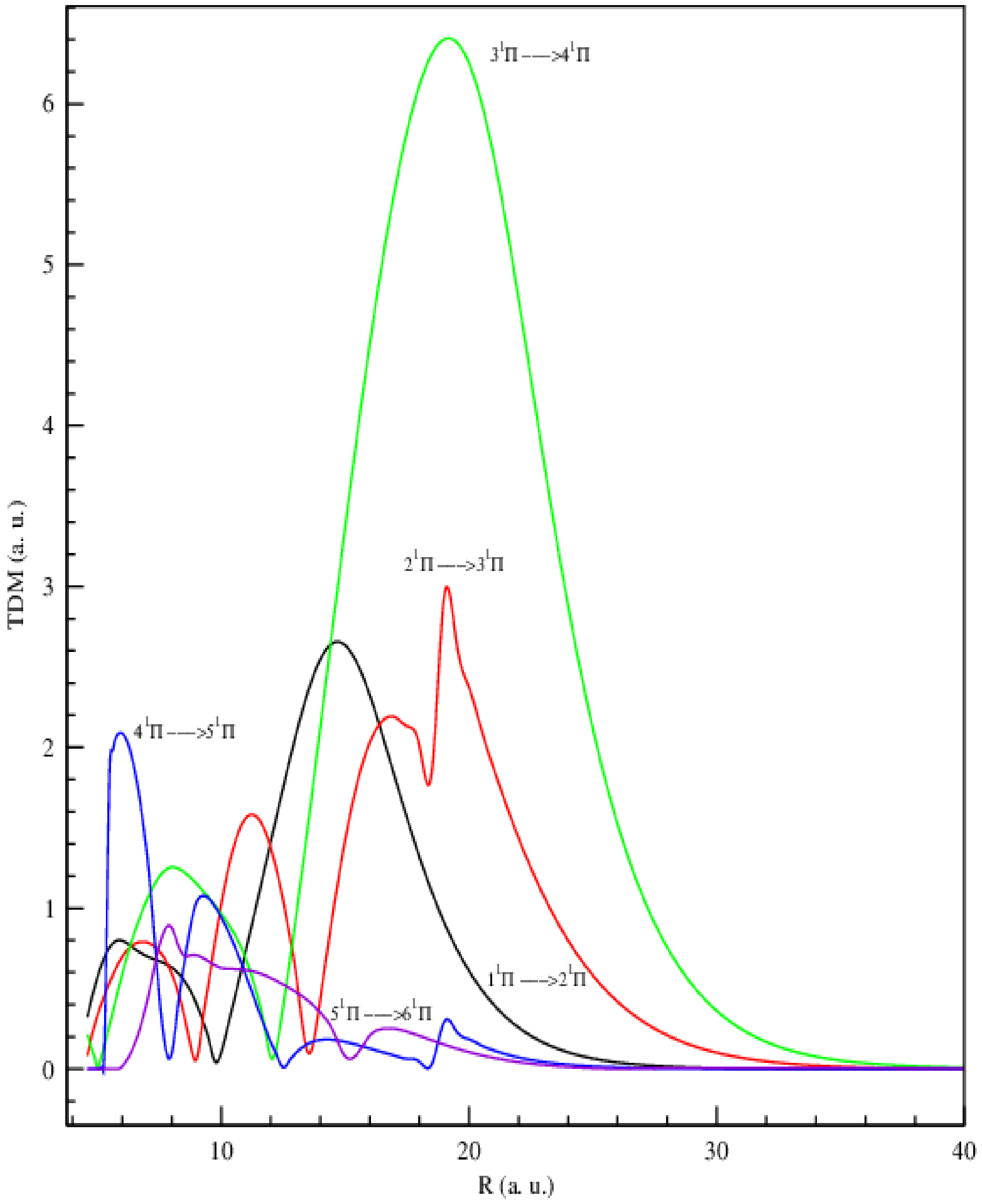

**Figure 15**

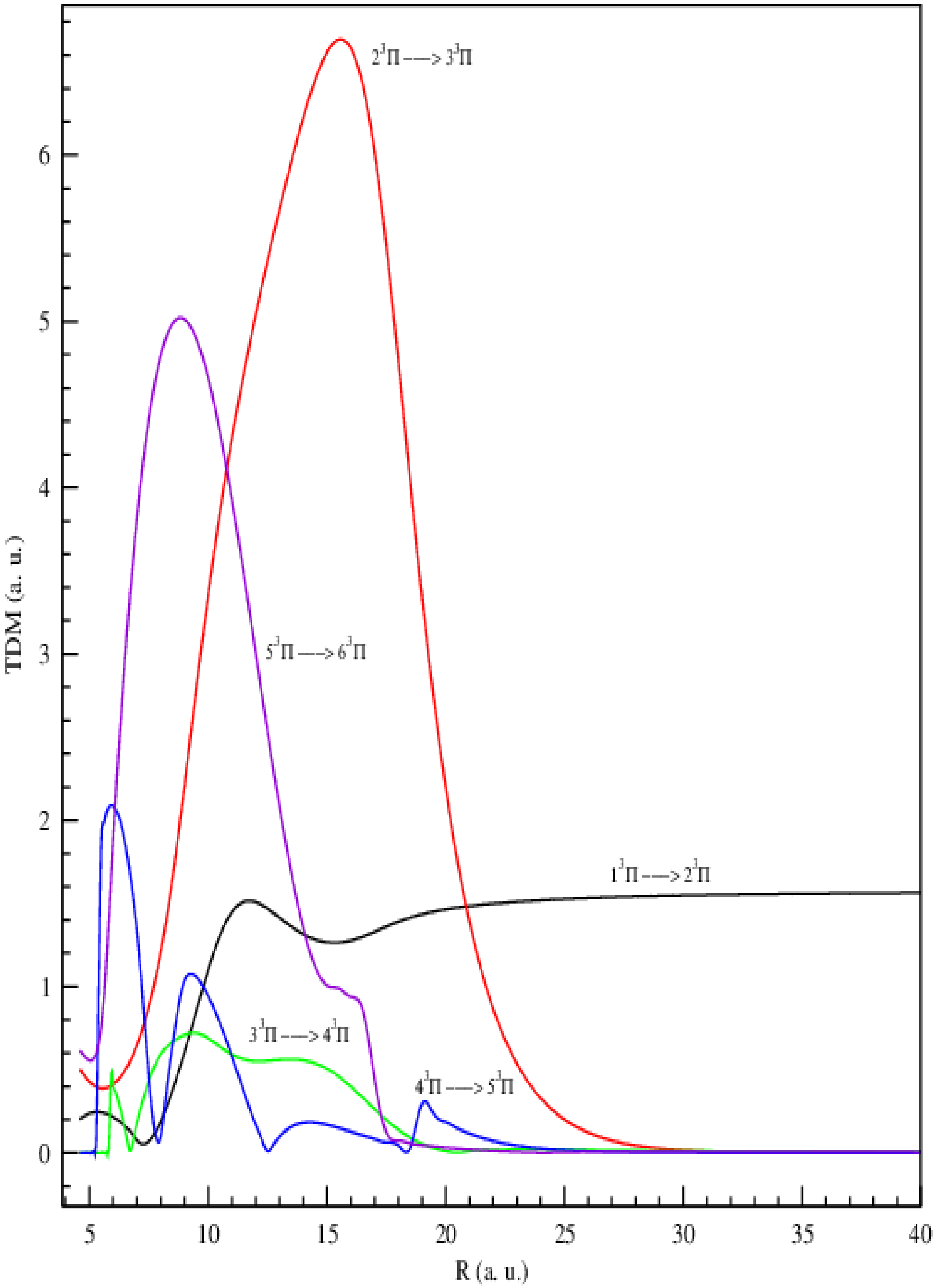

## Figure 16

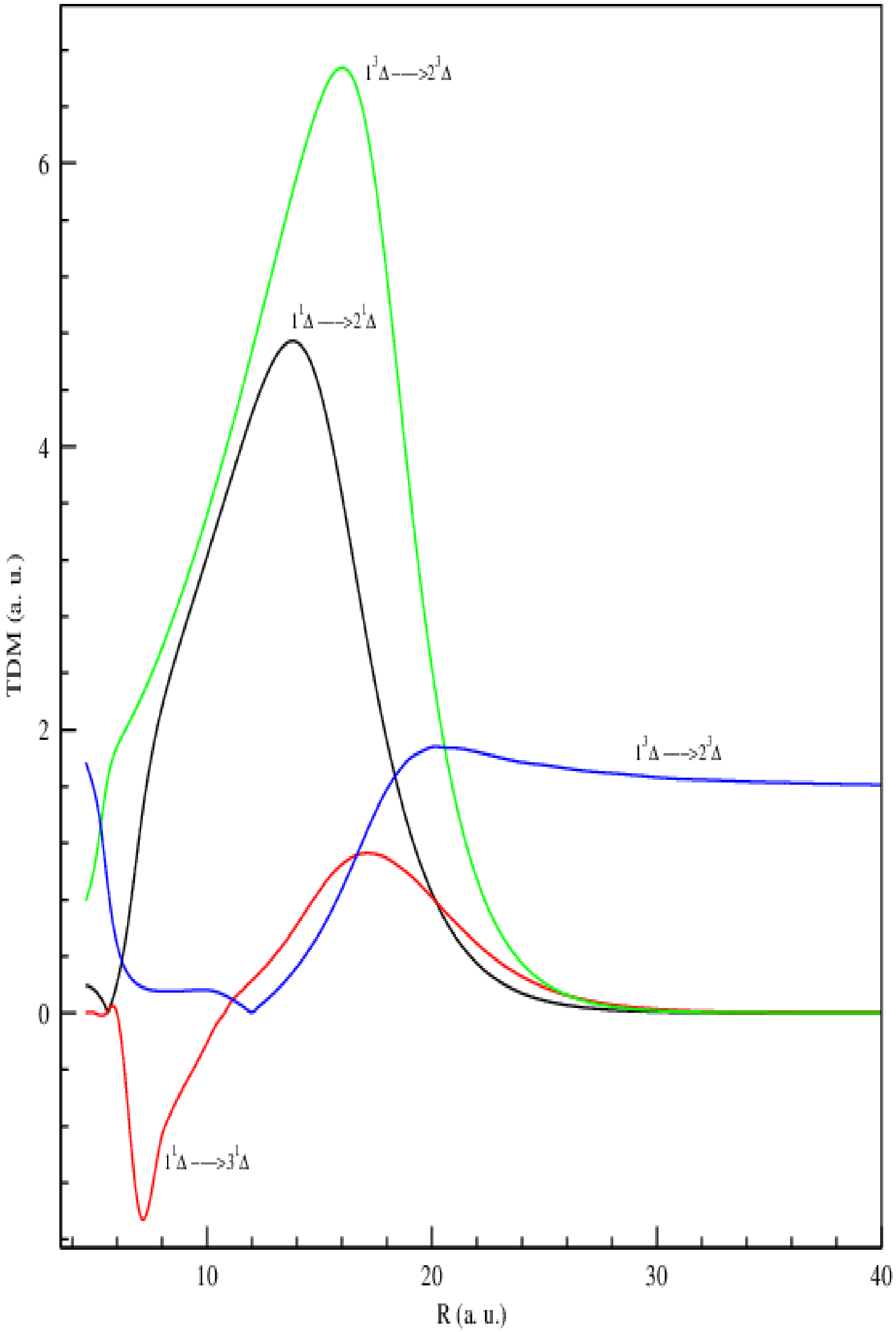

**Figure 17**

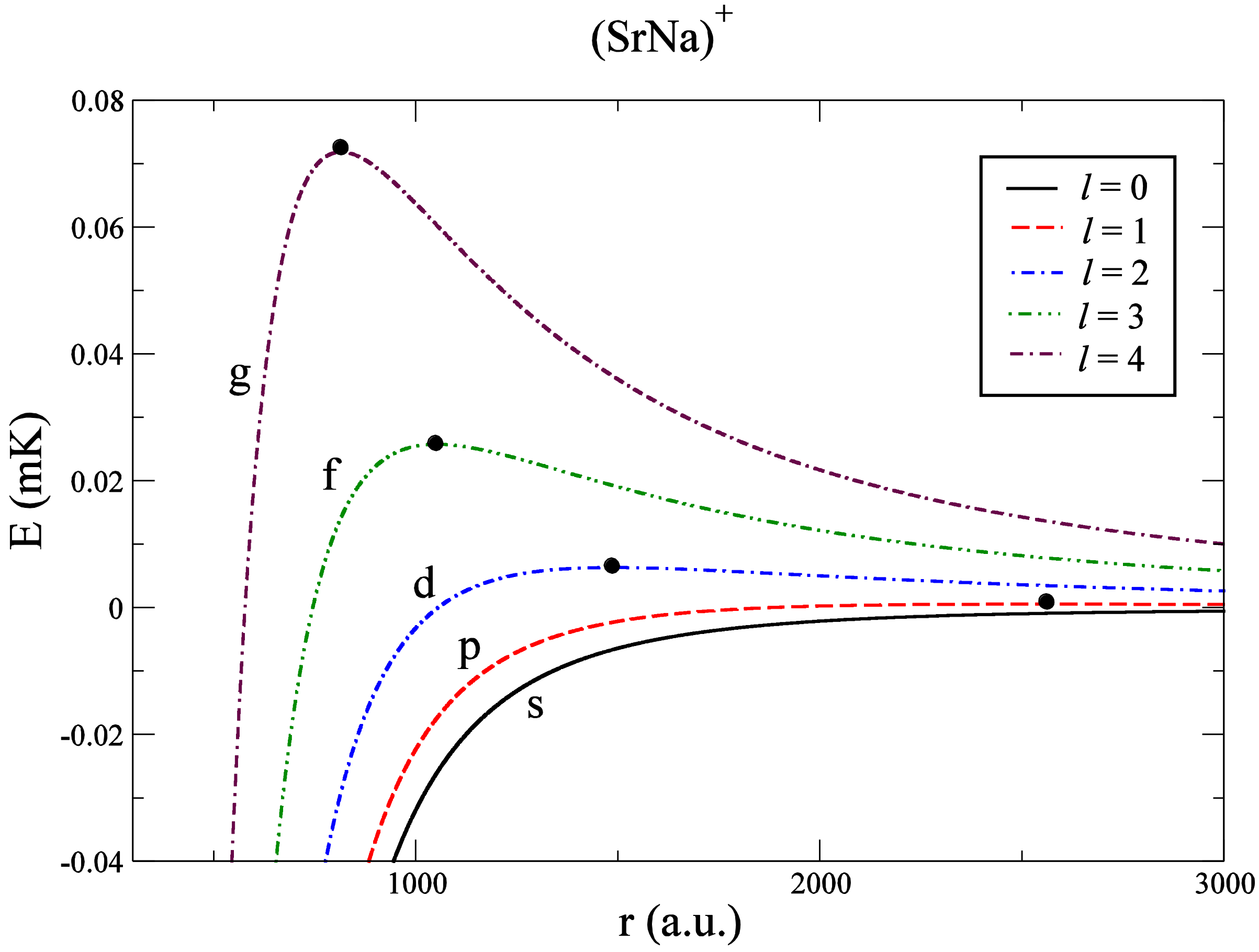

**Figure 18**

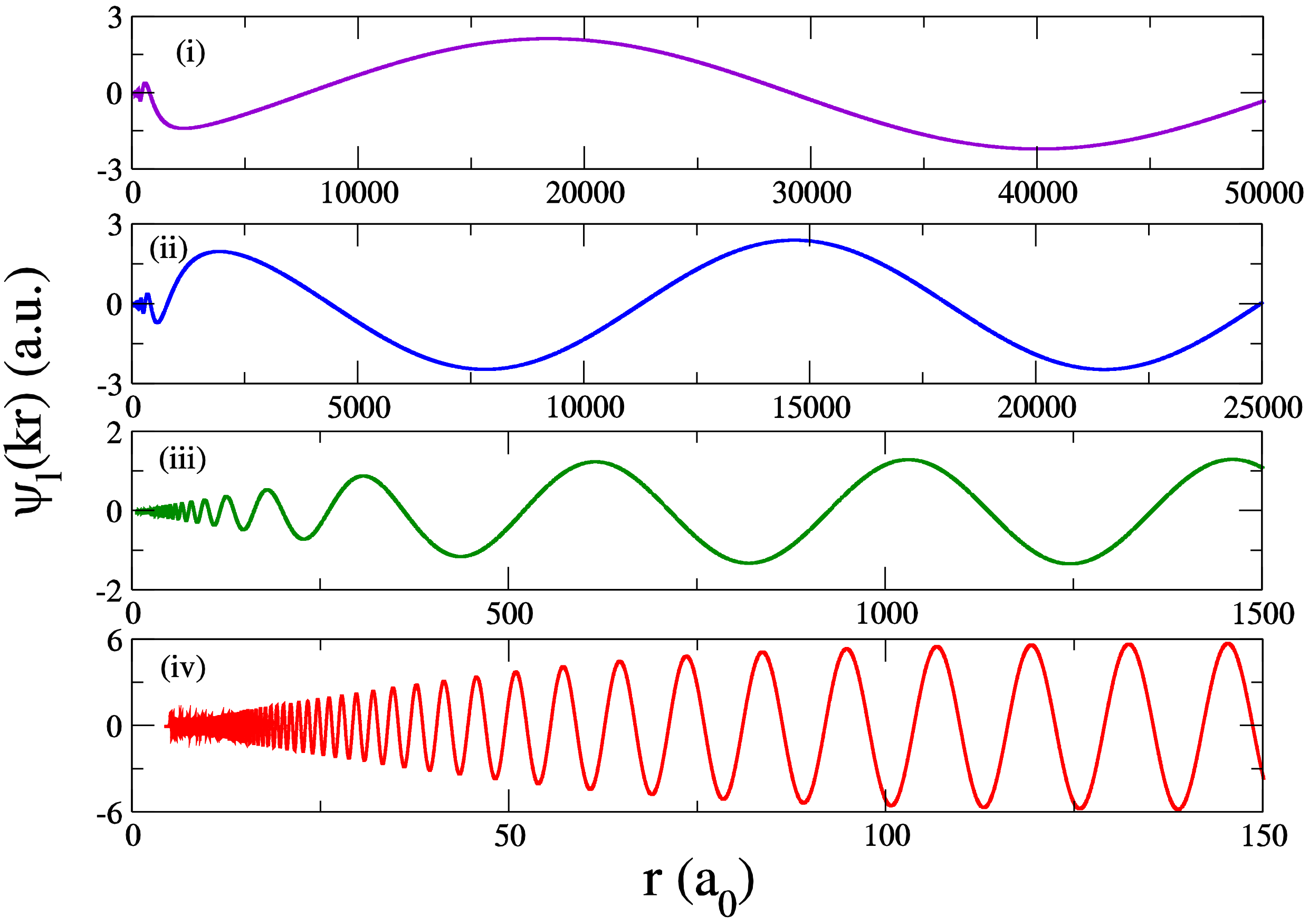

**Figure 19**

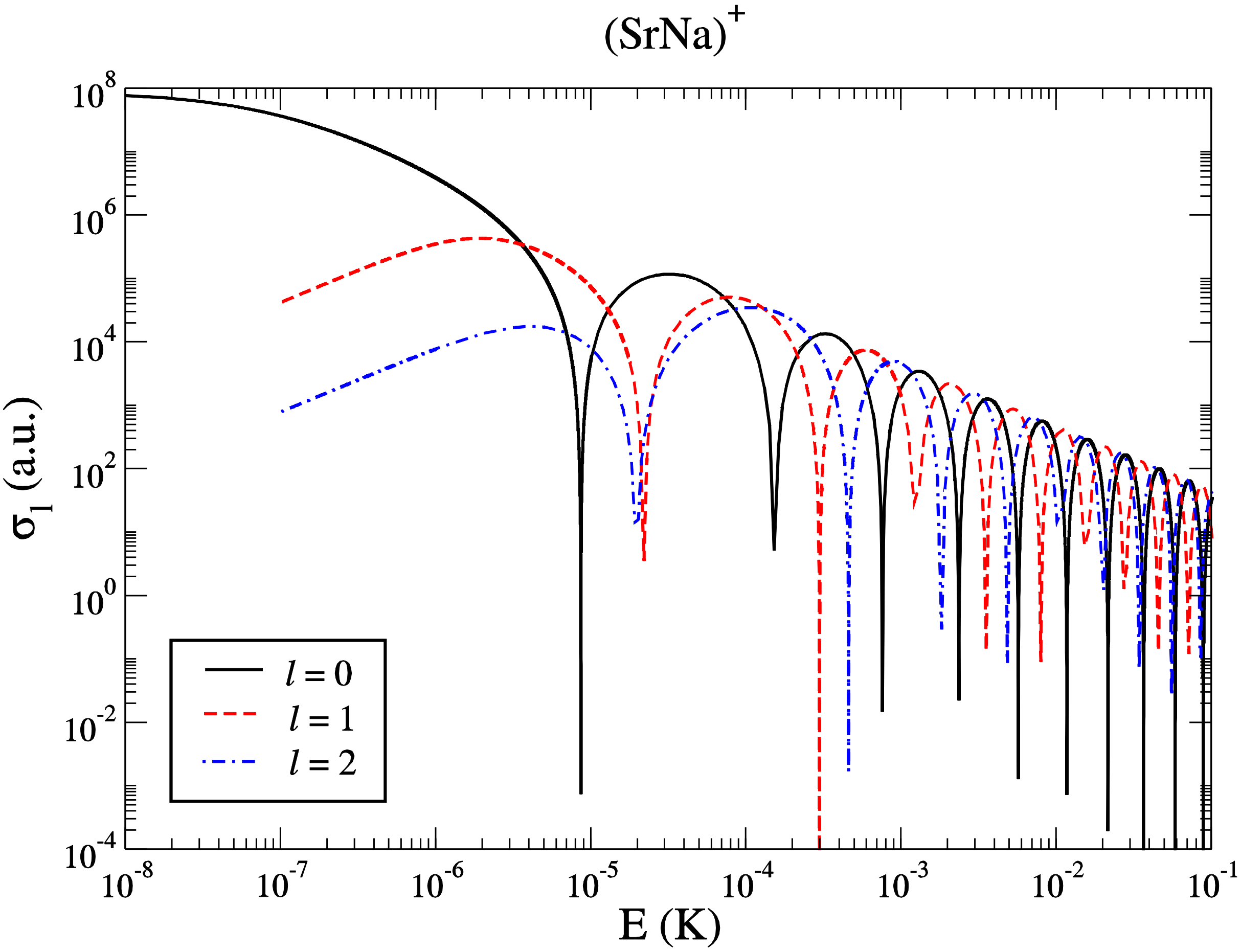

**Figure 20**

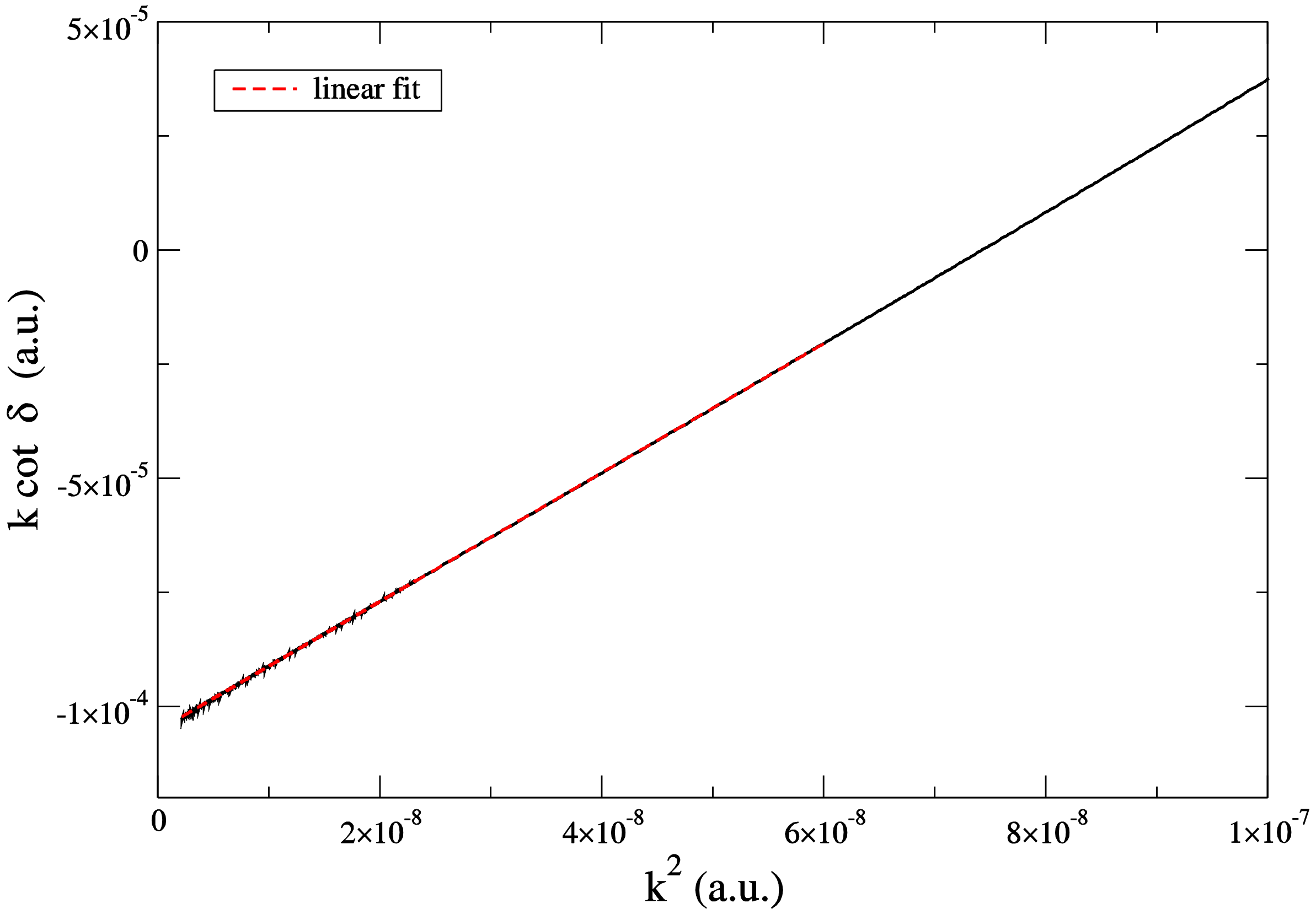

**Figure 21**

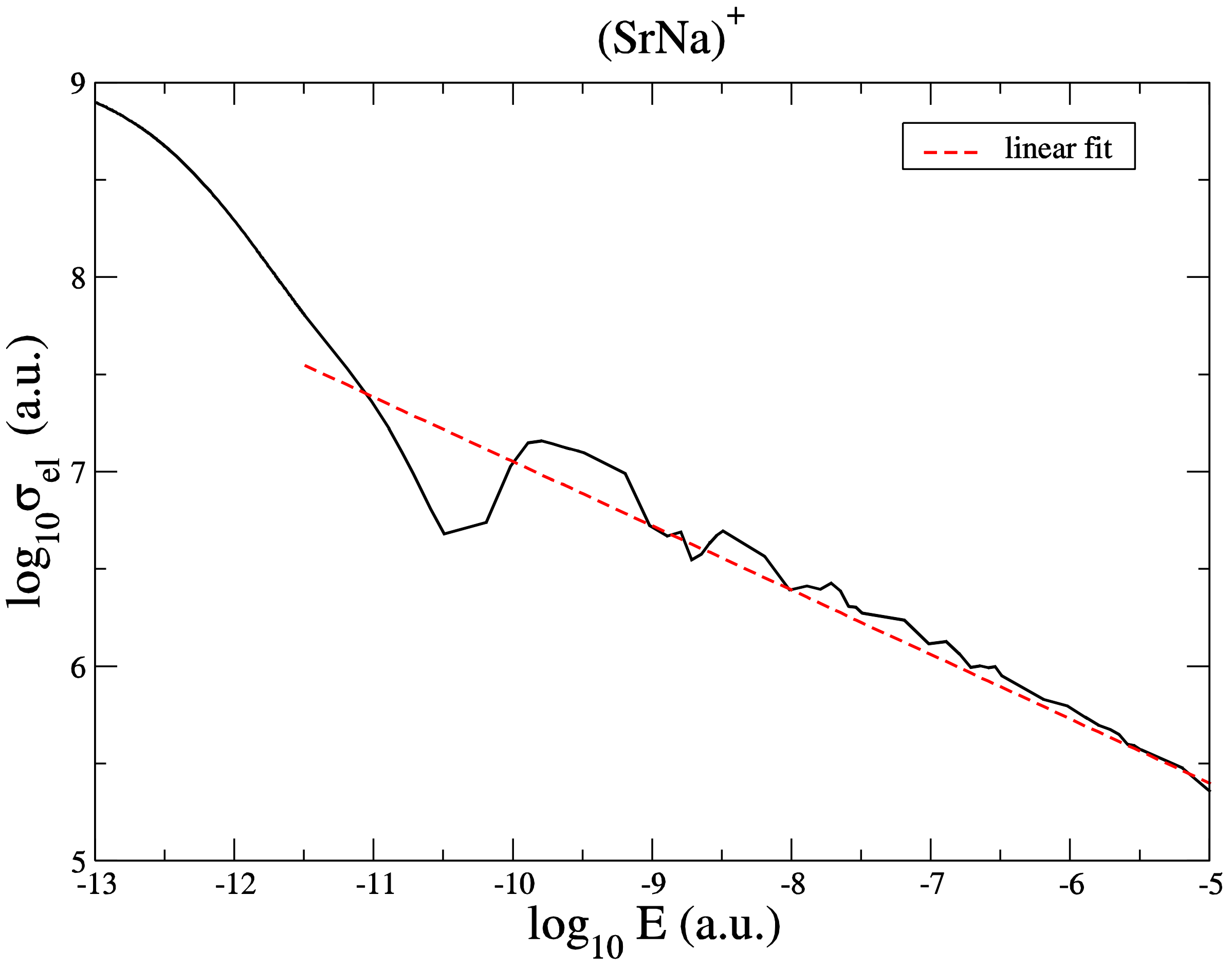